\def \VersionArXiv {}
\ifdefined\LaTeXdiff
	\let\VersionWithComments\undefined
	
\fi

\newcommand{\ourtitle}{Parametric schedulability analysis of a launcher flight control system under reactivity constraints}
\newcommand{\ourkeywords}{scheduling, real-time systems, model checking, parameter synthesis, \imitator{}}
\newcommand{\ouracks}{This work is partially supported by the ANR national research program PACS (ANR-14-CE28-0002)
	and
	ERATO HASUO Metamathematics for Systems Design Project (No.\ JPMJER1603), JST.}

\ifdefined\VersionArXiv
	\documentclass[]{article}
\else
	\documentclass[USenglish]{fundam}
	\usepackage{hyperref} %
\fi
\usepackage{graphicx}%

\usepackage[utf8]{inputenc}
\usepackage{float}

\usepackage[ruled,vlined,linesnumbered]{algorithm2e}
	\SetKwInOut{Input}{input}
	\SetKwInOut{Output}{output}

\usepackage{tabularx}
\usepackage{paralist} %
\usepackage{subfig}
\newenvironment{ienumerate}
	{\ifdefined\VersionArXiv\begin{enumerate}\else\begin{inparaenum}[\itshape i\upshape)]\fi}
	{\ifdefined\VersionArXiv\end{enumerate}\else\end{inparaenum}\fi}

\newenvironment{oneenumerate}
	{\ifdefined\VersionArXiv\begin{enumerate}\else\begin{inparaenum}[1)]\fi}
	{\ifdefined\VersionArXiv\end{enumerate}\else\end{inparaenum}\fi}

\usepackage{amsmath} %
\usepackage{amssymb} %
\usepackage{graphicx}
\graphicspath{{figures/}}

\usepackage[misc,geometry]{ifsym} %

\ifdefined\VersionArXiv
	\usepackage[backend=biber,backref=true,style=alphabetic,url=false,doi=true,defernumbers=true,sorting=anyt,maxnames=99]{biblatex} %
	\renewbibmacro*{doi+eprint+url}{%
		\iftoggle{bbx:doi}
			{\color{black!40}\footnotesize\printfield{doi}}
			{}%
		\newunit\newblock
		\iftoggle{bbx:eprint}
			{\usebibmacro{eprint}}
			{}%
		\newunit\newblock
		\iftoggle{bbx:url}
			{\usebibmacro{url+urldate}}
			{}%
	}

\fi
\ifdefined\VersionWithComments
	\usepackage{draftwatermark}
	\SetWatermarkText{draft}
	\SetWatermarkScale{3}
	\SetWatermarkColor[gray]{0.9}
\fi
\ifdefined\LaTeXdiff
	\usepackage{draftwatermark}
	\SetWatermarkText{diff}
	\SetWatermarkScale{3.5}
	\SetWatermarkColor[gray]{0.9}
\fi
\usepackage[svgnames,table]{xcolor}
\definecolor{darkblue}{rgb}{0, 0, 0.7}

\ifdefined\VersionArXiv
	\usepackage[
			pdfauthor={Etienne Andre, Emmanuel Coquard, Laurent Fribourg, Jawher Jerray and David Lesens},%
			pdftitle={\ourtitle{}},
			breaklinks  = true,
			colorlinks  = true,
			citecolor   = blue!50!blue,
			linkcolor   = darkblue,
			urlcolor    = blue!50!green,
		]{hyperref}
\fi

\usepackage[capitalise,english,nameinlink]{cleveref} %
\crefname{line}{\text{line}}{\text{lines}} %

\newcommand{\defProblem}[3]
{%
\noindent\fcolorbox{black}{blue!15}{
	\begin{minipage}{.95\columnwidth}
		\textbf{#1 problem:}\\
		\textsc{Input}: #2\\
		\textsc{Problem}: #3
	\end{minipage}
}
	
	\smallskip
	
}

\ifdefined \VersionArXiv
	\newcommand{\ArXivVersion}[1]{#1}
	\newcommand{\NOTArXivVersion}[1]{}
\else
	\newcommand{\ArXivVersion}[1]{}
	\newcommand{\NOTArXivVersion}[1]{#1}
\fi

\usepackage{tikz}
\usetikzlibrary{arrows,automata}
\tikzstyle{every node}=[initial text=,font=\footnotesize]
\tikzstyle{location}=[rectangle, rounded corners, minimum size=12pt, draw=black, fill=blue!10, inner sep=2pt]
\tikzstyle{invariant}=[draw=black, dotted, inner sep=1pt] %
\tikzstyle{urgent}=[fill=yellow, double]
\tikzstyle{bad}=[accepting,fill=red!50]

\usepackage{makecell}
\newcommand{\cellHeader}[1]{\cellcolor{blue!20}\textbf{#1}}
\newcommand{\cellNA}{\cellcolor{gray!50}N/A}

\newcommand{\cellYes}{\cellcolor{green!50}$\mathbf{\surd}$}
\newcommand{\cellNo}{\cellcolor{red!50}$\mathbf{\times}$}
\newcommand{\cellTO}{\cellcolor{red!50}TO}
\makeatletter
\def\renewtheorem#1{%
  \expandafter\let\csname#1\endcsname\relax
  \expandafter\let\csname c@#1\endcsname\relax
  \gdef\renewtheorem@envname{#1}
  \renewtheorem@secpar
}
\def\renewtheorem@secpar{\@ifnextchar[{\renewtheorem@numberedlike}{\renewtheorem@nonumberedlike}}
\def\renewtheorem@numberedlike[#1]#2{\newtheorem{\renewtheorem@envname}[#1]{#2}}
\def\renewtheorem@nonumberedlike#1{  
\def\renewtheorem@caption{#1}
\edef\renewtheorem@nowithin{\noexpand\newtheorem{\renewtheorem@envname}{\renewtheorem@caption}}
\renewtheorem@thirdpar
}
\def\renewtheorem@thirdpar{\@ifnextchar[{\renewtheorem@within}{\renewtheorem@nowithin}}
\def\renewtheorem@within[#1]{\renewtheorem@nowithin[#1]}
\makeatother

\renewtheorem{definition}{Definition}[section]
\renewtheorem{theorem}{Theorem}%
\renewtheorem{fact}{Fact}%
\renewtheorem{lemma}{Lemma}
\renewtheorem{example}{Example}
\renewtheorem{assumption}{Assumption}
\renewtheorem{proposition}{Proposition}
\renewtheorem{remark}{Remark}
\renewtheorem{corollary}{Corollary}
\renewtheorem{claim}[definition]{Claim}

\usepackage{listings}
\usepackage{color}

\definecolor{mygreen}{rgb}{0,0.6,0}
\definecolor{mygray}{rgb}{0.5,0.5,0.5}
\definecolor{mymauve}{rgb}{0.58,0,0.82}

\lstdefinestyle{ariane}{
	backgroundcolor=\color{white},   %
	basicstyle=\footnotesize,        %
	breakatwhitespace=false,         %
	breaklines=true,                 %
	captionpos=b,                    %
	commentstyle=\color{mygreen},    %
	deletekeywords={...},            %
	escapeinside={\%*}{*)},          %
	extendedchars=true,              %
	frame=single,	                   %
	keepspaces=true,                 %
	keywordstyle=\color{red!70!black}\bfseries,       %
	morekeywords={deadline,end,in,is,maf,ms,offset,out,period,processing,reactivity,thread,wcet,when},            %
	numbers=left,                    %
	numbersep=5pt,                   %
	numberstyle=\tiny\color{mygray}, %
	rulecolor=\color{black},         %
	showspaces=false,                %
	showstringspaces=false,          %
	showtabs=false,                  %
	stepnumber=1,                    %
	stringstyle=\color{mymauve},     %
	tabsize=2,	                   %
}

\definecolor{weborange}{RGB}{255,165,0}
\lstdefinestyle{imitator}{
	backgroundcolor=\color{white},   %
	basicstyle=\footnotesize,        %
	breakatwhitespace=false,         %
	breaklines=true,                 %
	captionpos=b,                    %
	commentstyle=\color{mygreen},    %
	deletekeywords={...},            %
	emph={True},
    emphstyle={\color{blue}\bfseries},
	escapeinside={\%*}{*)},          %
	extendedchars=true,              %
	frame=single,	                   %
	keepspaces=true,                 %
	keywordstyle=\color{red!70!black}\bfseries,       %
	language=Caml,                 %
	morekeywords={automaton,do,end,goto,invariant,loc,stop,sync,synclabs,urgent,when},            %
	numbers=left,                    %
	numbersep=5pt,                   %
	numberstyle=\tiny\color{mygray}, %
	rulecolor=\color{black},         %
	showspaces=false,                %
	showstringspaces=false,          %
	showtabs=false,                  %
	stepnumber=1,                    %
	stringstyle=\color{mymauve},     %
	tabsize=2,	                   %
	classoffset=1, %
	otherkeywords={>,<,-,+,&,:,=},
	morekeywords={>,<,-,+,&,:,=},
	keywordstyle=\color{weborange},
	classoffset=0,
}

\ifdefined\VersionWithComments
	\usepackage[colorinlistoftodos,textsize=footnotesize]{todonotes}
\else
	\usepackage[disable]{todonotes}
\fi
\newcommand{\gennote}[3]{\todo[linecolor=#2,backgroundcolor=#2!25,bordercolor=#2]{#3: #1}\xspace}
\newcommand{\ea}[1]{\gennote{#1}{blue}{ÉA}}

\ifdefined \VersionWithComments
	\newcommand{\todoinline}[1]{\mbox{}{\color{red}{\textbf{TODO}\ifx#1\\\else:\ \fi #1}}} %
\else
	\newcommand{\todoinline}[1]{}
\fi

\usepackage{verbatim} %

\ifdefined \VersionWithReview
	\usepackage{cancel}
	\usepackage{soul}
	\usepackage{marginnote}

	\newcommand{\reviewDelete}[1]{{\color{red}\st{#1}}}

\else

	\newcommand{\reviewDelete}[1]{}
\fi

\ifdefined\VersionArXiv
	\newcommand{\parORsubsection}{\subsection}
\else
	\newcommand{\parORsubsection}{\paragraph}
\fi

\newcommand{\init}{_0}

\newcommand{\A}{\ensuremath{\mathcal{A}}}

\newcommand{\Actions}{\Sigma}
\newcommand{\action}{\ensuremath{a}}
\newcommand{\C}{C}

\newcommand{\Clock}{\mathbb{X}} %
\newcommand{\ClockCard}{\ensuremath{{|\Clock|}}} %
\newcommand{\clock}{x} %
\newcommand{\clockval}{w} %
\newcommand{\ClocksZero}{\vec{0}}
\newcommand{\compOp}{\bowtie}

\newcommand{\edge}{e}
\newcommand{\Edges}{E}

\newcommand{\longueflecheRel}[1]{\stackrel{#1}{\mapsto}}

\newcommand{\flecheRel}{{\rightarrow}}

\newcommand{\grandn}{{\mathbb N}}
\newcommand{\grandq}{{\mathbb Q}}
\newcommand{\grandqplus}{\grandq_{+}} %
\newcommand{\grandr}{\ensuremath{\mathbb R}}
\newcommand{\grandrplus}{\ensuremath{\grandr_{+}}} %

\newcommand{\guard}{g}

\newcommand{\invariant}{\mathcal{I}}
\newcommand{\K}{\ensuremath{K}}

\newcommand{\loc}{\ensuremath{\ell}} %
\newcommand{\locinit}{\loc\init}
\newcommand{\Loc}{L} %

\newcommand{\Param}{\mathbb{P}} %
\newcommand{\param}{p} %
\newcommand{\ParamCard}{\ensuremath{{|\Param|}}} %
\newcommand{\pval}{v} %
\newcommand{\R}{{\mathbb{R}}}
\newcommand{\Rgeqzero}{\R_{\geq 0}}

\newcommand{\sinit}{s\init} %
\newcommand{\stopFunction}{\RTS}
\newcommand{\States}{S} %

\newcommand{\timelapseStop}[3]{#1^{\nearrow + #2}_{\setminus #3}}

\newcommand{\resets}{R}

\newcommand{\reset}[2]{\ensuremath{[#1]_{#2}}}
\newcommand{\valuate}[2]{\ensuremath{#2(#1)}}
\newcommand{\allActExcept}[1]{\ensuremath{\overline{#1}}}

\newcommand{\RTS}{\ensuremath{\textcolor{colorok}{\mathcal{S}}}}

\newcommand{\Processings}{\ensuremath{\textcolor{colorok}{\mathcal{P}}}}
\newcommand{\Reactivities}{\ensuremath{\textcolor{colorok}{\mathcal{R}}}}
\newcommand{\Threads}{\ensuremath{\textcolor{colorok}{\mathcal{T}}}}

\newcommand{\processing}{\ensuremath{\textcolor{colorok}{\mathbf{p}}}}
\newcommand{\reactivity}{\ensuremath{\textcolor{colorok}{\mathbf{r}}}}
\newcommand{\thread}{\ensuremath{\textcolor{colorok}{\mathbf{t}}}}

\newcommand{\procCont}{\ensuremath{\textcolor{colorok}{\processing_{\mathit{Cont}}}}}
\newcommand{\procGuid}{\ensuremath{\textcolor{colorok}{\processing_{\mathit{Guid}}}}}
\newcommand{\procMoni}{\ensuremath{\textcolor{colorok}{\processing_{\mathit{Moni}}}}}
\newcommand{\procNavi}{\ensuremath{\textcolor{colorok}{\processing_{\mathit{Navi}}}}}

\newcommand{\DR}{\ensuremath{\textcolor{colorok}{\mathit{DR}}}}
\newcommand{\DT}{\ensuremath{\textcolor{colorok}{\mathit{DT}}}}
\newcommand{\OT}{\ensuremath{\textcolor{colorok}{\mathit{OT}}}}
\newcommand{\MAF}{\ensuremath{\textcolor{colorok}{\mathit{MAF}}}}
\newcommand{\PP}{\ensuremath{\textcolor{colorok}{\mathit{PP}}}}
\newcommand{\PT}{\ensuremath{\textcolor{colorok}{\mathit{PT}}}}
\newcommand{\WCET}{\ensuremath{\textcolor{colorok}{\mathit{WCET}}}}

\newcommand{\WCETCont}{\ensuremath{\textcolor{colorok}{\WCET_{\mathit{Cont}}}}}
\newcommand{\WCETGuid}{\ensuremath{\textcolor{colorok}{\WCET_{\mathit{Guid}}}}}
\newcommand{\WCETMoni}{\ensuremath{\textcolor{colorok}{\WCET_{\mathit{Moni}}}}}
\newcommand{\WCETNavi}{\ensuremath{\textcolor{colorok}{\WCET_{\mathit{Navi}}}}}

\newcommand{\PPCont}{\ensuremath{\textcolor{colorok}{\PP_{\mathit{Cont}}}}}
\newcommand{\PPGuid}{\ensuremath{\textcolor{colorok}{\PP_{\mathit{Guid}}}}}
\newcommand{\PPMoni}{\ensuremath{\textcolor{colorok}{\PP_{\mathit{Moni}}}}}
\newcommand{\PPNavi}{\ensuremath{\textcolor{colorok}{\PP_{\mathit{Navi}}}}}

\newcommand{\styleSched}[1]{\ensuremath{\mathsf{#1}}}

\newcommand{\FPS}{\ensuremath{\styleSched{FPS}}}
\newcommand{\RMS}{\ensuremath{\styleSched{RMS}}}

\newcommand{\cheddar}{Cheddar}
\newcommand{\imitator}{\textsf{IMITATOR}}
\newcommand{\uppaal}{\textsc{Uppaal}}

\definecolor{loccolor1}{rgb}{.9, .95, 1}
\definecolor{loccolor2}{rgb}{.9, .95, 1}
\definecolor{loccolor3}{rgb}{.9, .95, 1}
\definecolor{loccolor4}{rgb}{.9, .95, 1}
\definecolor{loccolor5}{rgb}{1, .5, .5} %

	\definecolor{coloract}{rgb}{0.50, 0.70, 0.30}
	\definecolor{colorclock}{rgb}{0.4, 0.4, 1}
	\definecolor{colordisc}{rgb}{1, 0, 1}
	\definecolor{colorloc}{rgb}{0.4, 0.4, 0.65}
	\definecolor{colorparam}{rgb}{1, 0.6, 0.0}

\newcommand{\styleact}[1]{\ensuremath{\textcolor{coloract}{\mathrm{#1}}}}
\newcommand{\styleclock}[1]{\ensuremath{\textcolor{colorclock}{\mathit{#1}}}}

\newcommand{\styleloc}[1]{\ensuremath{\mathrm{#1}}}
\newcommand{\styleparam}[1]{\ensuremath{\textcolor{colorparam}{\mathrm{#1}}}}
\newcommand{\stylePTA}[1]{\ensuremath{\mathsf{#1}}}

\newcommand{\offsetTone}{\styleparam{offsetT1}}
\newcommand{\offsetTtwo}{\styleparam{offsetT2}}
\newcommand{\offsetTthree}{\styleparam{offsetT3}}
\newcommand{\deadlineTone}{\styleparam{deadlineT1}}
\newcommand{\deadlineTtwo}{\styleparam{deadlineT2}}
\newcommand{\deadlineTthree}{\styleparam{deadlineT3}}

\ifdefined \VersionWithComments
 	\definecolor{colorok}{RGB}{80,80,150}
\else
	\definecolor{colorok}{RGB}{0,0,0}
\fi

\newcommand{\eg}{\textcolor{colorok}{e.\,g.,}\xspace}
\newcommand{\ie}{\textcolor{colorok}{i.\,e.,}\xspace}

\newcommand{\wrt}{\textcolor{colorok}{w.r.t.}\xspace}
\makeatletter
\def\orcidID#1{\smash{\href{https://orcid.org/#1}{\protect\raisebox{-1.25pt}{\protect\includegraphics{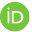}}}}}
\makeatother

\begin{document}

\ifdefined\VersionArXiv
\else
	\setcounter{page}{1001}
	\issue{XXI~(2001)}
\fi

\title{\ourtitle{}%
	\thanks{\ouracks{}}%
}

\ifdefined\VersionArXiv
\else
	\address{\url{eandre93430@lipn13.fr}}
\fi

\ifdefined\VersionArXiv
	\let\svthefootnote\thefootnote
	\newcommand\blankfootnote[1]{%
		\let\thefootnote\relax\footnotetext{#1}%
		\let\thefootnote\svthefootnote%
	}
	
	\begin{center}
		\textbf{\LARGE\ourtitle{}\blankfootnote{%
			This manuscript is the author version of the manuscript of the same name published in Fundamenta Informatica 182(1).
			This is an extended version of the manuscript published in the proceedings of the 19th International Conference on Application of Concurrency to System Design (ACSD 2019).
			The final authenticated version is available at \url{https://doi.org/10.3233/FI-2021-2065}.
			\ouracks{}
		}}
		
	\bigskip
	
		{\'Etienne Andr\'e$^{1,2,3}$\orcidID{0000-0001-8473-9555}, Emmanuel Coquard$^{4}$, Laurent Fribourg$^{5}$, Jawher~Jerray$^{6}$\orcidID{0000-0001-6170-7489} and David Lesens$^{4}$}
	\end{center}

	\bigskip

	\noindent{\footnotesize
		$^1$Université de Lorraine, CNRS, Inria, LORIA, Nancy, France\\
		$^2$JFLI, CNRS, Tokyo, Japan\\
		$^3$National Institute of Informatics, Tokyo, Japan\\
		$^4$ArianeGroup SAS, Les Mureaux, France\\
		$^5$Université Paris-Saclay, LSV, CNRS, ENS Paris-Saclay, France\\
		$^6$Université Sorbonne Paris-Nord, LIPN, CNRS, UMR 7030, F-93430, Villetaneuse, France\\
	}
	
\else
	\author{Étienne André \\
	Université de Lorraine, CNRS, Inria, LORIA, Nancy, France\\
	JFLI, CNRS, Tokyo, Japan\\
	National Institute of Informatics, Tokyo, Japan
	\and Emmanuel Coquard \\
	ArianeGroup SAS, Les Mureaux, France
	\and Laurent Fribourg \\
	Université Paris-Saclay, LSV, CNRS, ENS Paris-Saclay, France
	\and Jawher Jerray \\
	Université Sorbonne Paris-Nord, LIPN, CNRS, UMR 7030, F-93430, Villetaneuse, France 
	\and David Lesens \\
	ArianeGroup SAS, Les Mureaux, France
	}

	\maketitle

	\runninghead{É. André, E. Coquard, L. Fribourg, J. Jerray and D. Lesens}{Parametric schedulability of a flight control system}
\fi

\begin{abstract}
The next generation of space systems will have to achieve more and more complex missions.
	In order to master the development cost and duration of such systems, an alternative to a manual design is to automatically synthesize the main parameters of the system.
	In this paper, we present an approach for the specific case of the scheduling of the flight control of a space launcher.
	The approach requires two successive steps: (1) the formalization of the problem to be solved in a parametric formal model and (2) the synthesis of the model parameters with a tool.
	We first describe the problem of the scheduling of a launcher flight control, then we show how this problem can be formalized with parametric stopwatch automata; we then present the results computed by the parametric timed model checker \imitator{}.
	We enhance our model by taking into consideration the time for switching context, and we compare the results to those obtained by other tools classically used in scheduling.
\end{abstract}

\ifdefined\VersionArXiv
	\newcommand{\keywords}[1]
	{%
		\noindent\small\textbf{\textrm{Keywords---}} #1
	}


\else
\fi

\section{Introduction}\label{section:introduction}

Real-time systems combine concurrent behaviors with hard timing constraints.
An out-of-date reply is often considered as invalid even if its content is correct.
For \emph{critical} real-time systems, if a time constraint is violated, then the consequences may be disastrous.
Thus, a formal verification phase %
is essential in order to statically guarantee that all the tasks will be executed in their allocated time, and that the system will return results within the times guaranteed by the specification.

\ArXivVersion{%
Assessing the absence of timing constraints violations is even more important when the system can be hardly controlled once launched.
This is especially true in the aerospace area, where a system can only very hardly be modified or even rebooted after launching.
}

The next generation of space systems will have to achieve more and more complex missions.
In order to master the development cost and duration of such systems, an alternative to a manual design is to automatically synthesize the main parameters of the system.
While verifying a real-time system is already a notoriously difficult task, we tackle here the harder problem of \emph{synthesis}, \ie{} to automatically synthesize a part of the system so that it meets its specification.
We notably focus on the synthesis of admissible \emph{timing} values.

\parORsubsection{Contribution}
In this paper, we address the specific case of the scheduling of the flight control of a space launcher.
Our approach requires two successive steps:
\begin{ienumerate}
	\item the formalization of the problem to be solved in a parametric formal model and,
	\item the synthesis of the model parameters with a tool.
\end{ienumerate}
We first describe the problem of the scheduling of a launcher flight control;
then we formalize this %
	problem
with parametric stopwatch automata, an extension of timed automata~\cite{AD94} with parameters%
~\cite{AHV93} and stopwatches (``the ability to stop clocks'')~\cite{CL00,SSLAF13};
third, we present the results computed by the \imitator{}~\cite{Andre21} tool.
We compare our results with those obtained by other tools classically used in scheduling.
A key aspect is the verification and synthesis under some \emph{reactivity constraints}: the time from a data generation %
by sensor's measurement which is considered as an input
to its output must always be less than a threshold.
The solution we propose is compositional, in the sense that the reactivity constraints can be checked independently.
We consider both an instantaneous switch from one thread to another, and more general systems where the switch between two threads has a CPU cost due to the copy of data between the contexts of each thread.

We propose here a solution to the problems using an extension of parametric timed automata (PTA), which are an extension of finite state automata with clocks and parameters~\cite{AHV93}.
The general class of PTA is notoriously undecidable, and notably the mere problem of deciding whether at least one parameter valuation allows the system to reach a global state is undecidable, even over discrete time~\cite{AHV93}, with a single integer-valued parameter over dense time~\cite{BBLS15}, or with a single clock compared to parameters~\cite{Miller00} (see~\cite{Andre19STTT} for a survey).
Still, some decidable subclasses were proposed (\eg{} \cite{HRSV02,BlT09,ALime17,ALR18FORMATS}), notably in the field of scheduling real-time systems~\cite{CPR08,Andre17FMICS}.
In spite of these undecidability results, the use of parametric timed automata for solving various concrete problems was recently considered, in frameworks such as hardware verification~\cite{CEFX09}, analysis of music scores~\cite{FJ13}, monitoring~\cite{AHW18} or software product lines testing~\cite{LGSBL19}.
We show here that this formalism is also useful for solving concrete scheduling problems---such as the one considered here.

\parORsubsection{Outline}
After discussing related works in \cref{section:related},
\cref{section:problem} presents the problem we aim at solving.
\cref{section:PSA} recalls parametric stopwatch automata.
\cref{section:specification} exposes our modeling;
we extend our solution in \cref{section:compositional} in a compositional manner, and in \cref{section:switch} to enhance the model with context switch times.
\cref{section:experiments} gives the results obtained, while 
\cref{section:comparison} makes a comparison with other tools of the literature (solving only a part of the problem).
\cref{section:conclusion} concludes the paper.

\section{Related works}\label{section:related}
\parORsubsection{Scheduling}
A long line of works in the last five decades has been devoted to the problem of scheduling analysis of real-time systems with various flavors.
Several analytical methods were proposed to study the schedulability for a particular situation.
Such analytical methods need to be tuned for each precise setting (uniprocessor or multiprocessor, scheduling policy, absence or presence of offsets, jitters, etc.).
Most of them do not cope well with uncertainty.
For example, in~\cite{BB97}, three methods for the schedulability analysis with offsets are proposed.
In~\cite{BB04}, an efficient approach for testing schedulability for \RMS{} (rate monotonic) in the case of (uniprocessor) analysis is proposed, through a ``parameter'' (different from our timing parameters) to balance complexity versus acceptance ratio.
\parORsubsection{Scheduling with model checking}

Schedulability with model checking is a trend that started as early as the first works on timed model checking (\eg{} \cite{WME92,AHV93,AD94,YMW97,CC99}), and grew larger since\ArXivVersion{ the early 2000s}.

A natural model to perform schedulability analysis is (extensions of) timed automata (TA)~\cite{AD94}.
On the negative side, the cost of state space explosion often prevents the verification of very large real-time systems.
On the positive side, they allow for more freedom, and can model almost any system with arbitrarily complex constraints;
in addition, despite the cost of state space explosion, they can be used to verify small to medium-size systems for which no other method is known to apply.

In \cite{AM01,AM02}, (acyclic) TA are used to solve job-shop problems.
The preemption is encoded in~\cite{AM02} with \emph{stopwatches}, while keeping some decidability results.
In~\cite{AAM06}, scheduling is performed using TA.
Timed automata allow to model naturally and verify more complex systems, which are not captured so easily in traditional formalisms for schedulability analysis.

In \cite{NWY99,FKPY07}, task automata are proposed as a formalism extending TA to ease the modeling (and the verification) of uniprocessor real-time systems: in some cases, the schedulability problem of real-time systems is transformed into a reachability problem for standard TA and it is thus decidable.
This allows %
applying
model-checking tools for TA to schedulability analysis with several types of tasks and most types of scheduler.

In \cite{SLSFM14}, hierarchical scheduling systems are encoded using linear hybrid automata, a model that generalizes TA.
This approach outperforms analytical methods in terms of resource utilization.
In~\cite{SL14}, linear hybrid automata are used to perform schedulability analysis for multiprocessor systems under a global fixed priority scheduler: this method is more scalable than existing exact methods, and shows that analytical methods are too pessimistic.

In~\cite{FLSC16}, a schedulability analysis method is introduced using the model of \emph{timed regular task automata} (using under-approximated WCETs) and then using nested timed automata; this method is shown to be exact.
The problem we solve here shares similarities with analyses done in~\cite{FBGLP10,MLRNSPPH10}.
An important difference between \cite{FBGLP10,MLRNSPPH10} and our case study comes from the fact that, here, there are two distinct notions of ``thread'' and ``processing'', while in~\cite{FBGLP10,MLRNSPPH10} there was only one notion called ``task''.
Most importantly, none of these works consider timing parameters.
\parORsubsection{Scheduling with parameters}
When some of the design parameters are unknown or imprecise, the analysis becomes much harder.
Model checking with parameters can help to address this.
In~\cite{CPR08}, PTA are used to encode real-time systems so as to perform parametric schedulability analysis.
A subclass (with bounded offsets, parametric WCETs but constants deadlines and periods) is exhibited that gives exact results.
In contrast, our work allows for parameterized deadlines; in addition, reactivities are not considered in~\cite{CPR08}.

In~\cite{FLMS12}, we performed robust schedulability analysis on an industrial case study, using the inverse method for PTA~\cite{ACEF09,AS13} implemented in \imitator{}.
While the goal is in essence similar to the one in this manuscript, the system differs: \cite{FLMS12} considers multiprocessor, and preemption can only be done at fixed instants, which therefore resembles more Round Robin than real \FPS{}.
In~\cite{SSLAF13}, we showed that PTA-based methods are significantly more complete than existing analytical methods to handle uncertainty.
In~\cite{SAL15}, we solved an industrial challenge by Thales using \imitator{}; in~\cite{AFMS19}, we verified an industrial asynchronous leader election algorithm using \imitator{}, with additional abstractions.

In \cite{LPPR13}, the analysis is not strictly parametric, but concrete values are iterated so as to perform a cartography of the schedulability regions.
However, the resulting analysis of the system is incomplete.

In \cite{BHJL16}, timed automata are extended with multi-level clocks, of which exactly one at a time is active.
The model enjoys decidability results, even when extended with polynomials and parameters, but it remains unclear whether concrete classes of real-time systems can actually be modeled.

The aforementioned task automata were extended in~\cite{Andre17FMICS} to \emph{parametric} task automata; some schedulability problems remain decidable in this setting, \ie{} it is possible in some cases to decide whether the set of valuations ensuring schedulability is empty or not.
In addition, procedures for exhibiting schedulability regions are proposed and implemented.

Finally, \textsc{Roméo}~\cite{LRST09} also allows for parametric schedulability analysis using parametric time Petri nets~\cite{TLR09}, with applications to critical real-time systems~\cite{PRHSRLA16}.

\section{Description of the system and problem}\label{section:problem}

The flight control of a space launcher is classically composed of three algorithms:
\begin{ienumerate}
	\item The \emph{navigation} computes the current position of the launcher from the sensor's measurement (such as inertial sensors);
	\item The \emph{guidance} computes the most optimized trajectory from the launch pad to the payload release location;
	\item The \emph{control} orientates the thruster to follow the computed trajectory.
\end{ienumerate}
Due to the natural instability of a space launcher, strict real-time requirements have to be respected by the implementation of the flight control: frequency of each algorithm and reactivity between the sensor's measurement acquisition and the thruster's command's sending.

The case study described in this paper is a  simplified version of a flight control composed of a navigation, a guidance, a control and a monitoring algorithms; these four parts are called \emph{processings}\footnote{%
	Following the vocabulary used within ArianeGroup.
	Technically, a processing is a \emph{node} in SCADE, \ie{} a subprogram activated cyclically, with a frequency, an activation condition and inputs/outputs.
} in the following.
Each processing has a name and a required rational-valued period; in our setting, the processing deadline is equal to the period.
A processing can potentially read data from the avionics bus (``in'' data) and/or write data to the same avionics bus (``out'' data).
\cref{figure:example_control_system} shows an example of such a system (all the numerical data provided in this paper are only examples that do not necessarily correspond to an actual system).

\begin{figure*}[tb]
	\centering
\begin{lstlisting}[style=ariane]
processing Navigation (Meas     : in)  is period (5ms);  end;
processing Guidance                    is period (60ms); end;
processing Control    (Cmd      : out) is period (10ms); end;
processing Monitoring (Safeguard: out) is period (20ms); end;
\end{lstlisting}
	\caption{An example of a flight control system}
	\label{figure:example_control_system}
\end{figure*}
\subsection{Threads and deterministic communications}\label{ss:threads}

The software components of the system are physically deployed on a single processor~\cite{OGL06}.
Processings are allocated on \emph{threads} run by the processor.
\begin{figure}[htb]
	\centering
  \includegraphics[width=.8\columnwidth]{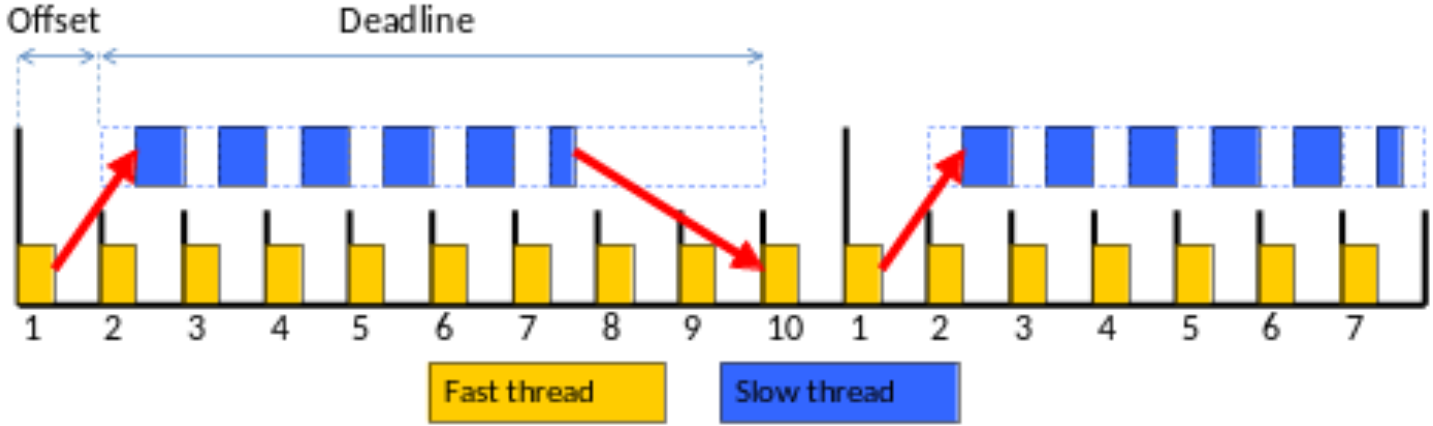}
  \caption{The communication between threads}
  \label{fig:communication_threads}
\end{figure}

\cref{fig:communication_threads} exemplifies the way data are exchanged between two threads.
The fast thread (in yellow) has a period of~1 time unit.
This period defines the time granularity of the system (this implies that the offset of the fast thread is 0 and that its deadline is~1).
In this example, the slow thread (in blue) has an offset of~1 (its start is delayed by 1 cycle compared to the start of the fast thread), a period of~10 and a deadline of~8.
The numbers from 1 to~10 denote the index of the period of the fast thread within the period of the slow thread%
.\label{newtext:fig2}
The first communication between the fast thread and the slow one is performed at the end of the first \emph{period}; this explains that, although the second occurrence of the fast thread finishes before the first occurrence of the slow thread, this is the \emph{first} occurrence of the fast thread which is communicated to the slow thread.
Similarly, in order to ensure the determinism and taking into the priority between the threads, the communication between the slow thread and the fast thread is performed at the \emph{deadline} of the slow thread, \ie{} at the end of cycle 9 (offset + deadline).
That is, the first occurrence of the fast thread to receive data from the slow thread is not the one starting at $t=8$ nor at $t=9$, but the one starting at $t=10$.

In our setting, all the thread periods are \emph{harmonic}, \ie{} a thread period is a multiple of the period of the thread just faster (they pairwise divide each other). %
In other terms, for a system that contains a set of threads $\thread_1,\ldots , \thread_k$, all the thread periods are considered harmonic 
if for every thread~$\thread_j$ (for all $j \in \{1,\ldots ,k\}$), the period~$\PT_j$ of~$\thread_j$ is a multiple of the periods %
of all the threads of smaller period, that is, $\PT_j$ is a multiple of the periods of all threads $\thread_i \in \{\thread_1,\ldots ,\thread_k \mid \PT_i <\PT_j \}$.
In our case, the harmonic assumption on threads will not affect the modeling of the system in \cref{section:specification}; however, it may be used to reduce the number of clocks and considerably decrease the computation time of our approach.

In addition, in order to ensure the determinism of the scheduling (which facilitates the verification of the system), the threads work in a synchronous manner:
\begin{itemize}
	\item The inputs of a thread are read \emph{at its start}; that is, no inputs are read during the execution of the thread.
	\item The outputs of a thread are provided \emph{at its deadline}; that is, not only no outputs are provided during the execution of the thread, but the output is also not provided as soon as the thread terminates---if it terminates before its deadline---but only at its deadline.
\end{itemize}

\paragraph{Switch time}
In our case study, the \emph{switch} time, \ie{} the time needed by the CPU to copy memory information when changing threads, is 500\,$\mu s$.

\subsection{Reactivities}\label{ss:reactivities}

To ensure the controllability of the launcher, a \emph{reactivity}\footnote{%
	In the literature, the term ``reactivity'' is also referred to as ``latency'' (see, \eg{} \cite{FBGLP10}).\label{footnote:latency} %
} is required between a data read from the avionics bus (a measurement) and a data written to the avionics bus (a command).
A reactivity imposes a maximum bound on the time required by these data to ``travel'' from the measurement to the command.
This concept is quite similar to that studied in~\cite{FBGLP10} (without timing parameters).

\begin{definition}[reactivity]\label{definition:reactivity}
	A \emph{reactivity constraint} imposes an upper bound from a data read from the avionics bus to a data written to the avionics bus, where the sequence of the path of the data represents a precedence constraint.
\end{definition}

Several paths are potentially possible between a read data and a written data.
\cref{figure:reactivities} shows an example of such reactivities.
\begin{figure}[htbp!]
  \centering

\begin{lstlisting}[style=ariane]
reactivity Meas -> Navigation -> Guidance  -> Control -> Cmd       is 150ms;
reactivity Meas -> Navigation -> Control              -> Cmd       is 15ms;
reactivity Meas -> Navigation -> Monitoring           -> Safeguard is 55ms;
\end{lstlisting}
  \caption{Some typical reactivities}
  \label{figure:reactivities}
\end{figure}

\begin{figure*}[htbp!]
  \centering
  
  \includegraphics[width=.8\textwidth]{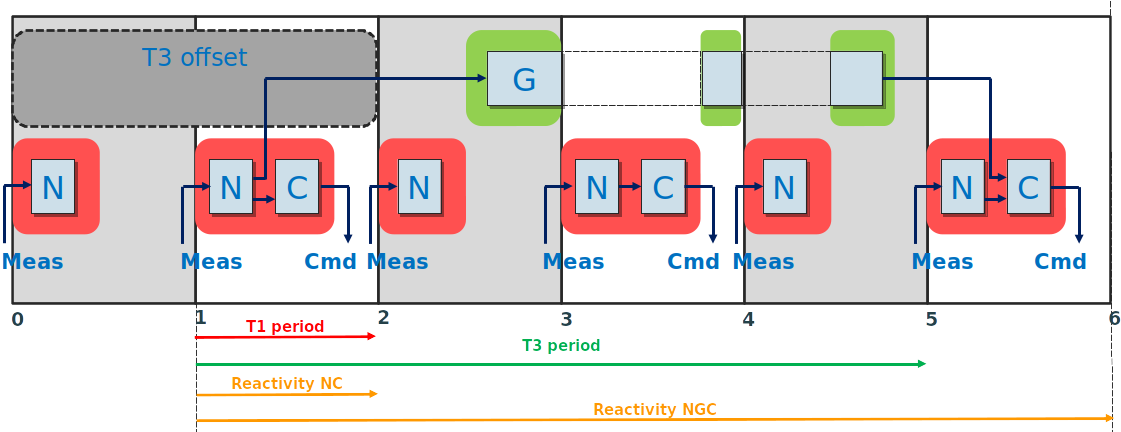}
  \caption{Determinism and reactivities}
  \label{figure:example-reactivities}
\end{figure*}

Reactivities too must follow the deterministic communication model from \cref{ss:threads}.
Consider the execution of threads and processings in \cref{figure:example-reactivities} (the values of periods and WCETs are given for illustration purpose, and do not correspond to the ones from our case study).
Consider the reactivity imposing that the sequence of data ``Meas $\rightarrow$ Navigation $\rightarrow$ Guidance $\rightarrow$ Control $\rightarrow$ Cmd'' should be equal to or less than~5.
Due to the data being communicated at the end of each thread only, the Guidance processing (marked with green ``G'' in \cref{figure:example-reactivities}) does not receive the data from the third execution of the Navigation processing (marked with ``N'' in red), as the data of the third Navigation will be sent at the end of the thread T1 period, but from the second execution of Navigation.
Therefore, in \cref{figure:example-reactivities}, the only path of interest is the path of the data starting from the second execution of Meas, going to the second execution of Navigation, then going to the (only) execution of Guidance, and then finishing in the third execution of Control, before being written to the third occurrence of Cmd.
Also note that the data output by the first execution of Navigation are successfully sent to T2 at the end of the first period of~T1, but will be overwritten by the second occurrence of Navigation, and are therefore not of interest when checking the satisfaction of reactivities.
Therefore, the time from the production of these data (at $t = 1$) to their writing on the avionic bus (at $t = 6$) is~5, and therefore the reactivity is satisfied.

We want to solve the scheduling problem of periodic processings \emph{under reactivity constraints}.

\subsection{Processings and assignment into threads}

A WCET (worst case execution time) is measured or computed for each processing.
An example is given in \cref{figure:example_WCET}.

\begin{figure}[htbp!]
  \centering

\begin{lstlisting}[style=ariane]
processing wcet Navigation (1ms);
processing wcet Guidance   (15ms);
processing wcet Control    (3ms);
processing wcet Monitoring (5ms);
\end{lstlisting}
  \caption{Example of Worst Case Execution Times}
  \label{figure:example_WCET}
\end{figure}

An important problem is to find a proper assignment of the processings into threads, with their respective periods, while minimizing the number of threads.
A solution to this problem consists of a set of cyclic threads on which the processings are deployed.
In our setting, these threads %
are scheduled with a preemptive and fixed priority policy (\FPS{}). %
A thread has a name and is defined by the following data:
\begin{enumerate}
	\item a rational-valued period;
	\item a rational-valued offset (with 0 $\leq$ offset $<$ period), \ie{} the time from the system start until the first periodic activation;
	\item a rational-valued (relative) deadline (with 0 $<$ deadline $\leq$ period), \ie{} the time after each thread activation within which all processings of the current thread should be completed;
	\item a rational-valued major frame (or ``MAF''). A MAF defines the duration of a pattern of processing activation.\ea{non défini !}
	The MAF in this case study is equal to~10.
	\item a set of processings deployed on the thread. 
		Different processings may be executed %
			in an order which may change %
			at each cycle.\label{newtext:diffproc}
	 However, after a MAF duration, the same pattern of processings is repeated.
\end{enumerate}

In order to simplify the scheduling problem, we have considered in this paper a pre-allocation of processings on threads, as specified in \cref{figure:typical_solution}: that is, Navigation and Control are allocated on T1, while Monitoring and Guidance are allocated on T2 and~T3, respectively.
In addition, Navigation is executed at every period of T1, while Control is executed (after Navigation) on \emph{odd} cycles only; this is denoted by the \texttt{when 1} syntax in \cref{figure:typical_solution}.

\subsection{A formal framework for real-time systems}

A real-time system $\RTS = \{\Processings, \Threads, \Reactivities\}$ is viewed here as
	a set of \emph{processings}
		$\Processings= \{\processing_1 , \processing_2 , \cdots , \processing_{m} \}$,
	a set of \emph{threads} $\Threads = \{\thread_{1} , \thread_{2} , \cdots , \thread_{n} \}$
	and a set of \emph{reactivities} $\Reactivities= \{\reactivity_{1} , \reactivity_{2} , \cdots, \reactivity_{q} \}$.
A thread $\thread_{i}$ computes a usually infinite stream of processings instances. 

In our setting, a thread $\thread_{i}$ is periodic, \ie{} generates instances every fixed amount of time (the ``period''), and is characterized by a 5-tuple $(\PT_{i} , \OT_{i} , \DT_{i}, \MAF_{i}, \Processings_{i})$, where $\PT_{i}$ corresponds to the period, $\OT_{i}$ to the offset, $\DT_{i}$ to the deadline, $\MAF_{i}$ defines the duration of a pattern of processing activation $\processing_{i_k}$ (where $\processing_{i_k}$ denotes the $k$th proccessing computed in thread~$\thread_{i}$)%
	, and $\Processings_{i}$ defines a subset of processings of $\Processings$ allocated to $\thread_{i}$.
\footnote{%
	Note that the MAF is a per-thread property; it is quite similar to the ARINC 653 standard used in industrial civil airplane~\cite{CGN13} designs, except that the ARINC 653 is a per-CPU property.\label{newtext:MAF}
}

A processing $\processing_i$ is characterized by two values $\WCET_{i}$ (Worst Case Execution Time) and~$\PP_{i}$ (Processing Period): When a processing is activated, it is executed for at most time $\WCET_{i}$ time units every~$\PP_{i}$ time units.

\begin{figure}[htbp!]
  \centering
  
	\ifdefined\VersionArXiv
		\newcommand{\scalefactorRTchar}{.75}
	\else
		\newcommand{\scalefactorRTchar}{.85}
	\fi
  \usetikzlibrary{decorations.pathreplacing,positioning, arrows.meta}
\usetikzlibrary{positioning}
\usetikzlibrary{arrows}

\begin{center}
\scalebox{\scalefactorRTchar}{
\begin{tikzpicture}[node distance=1mm]
\coordinate (d) at (0,0) node[below=of d] {0};
\coordinate (e) at (15.5,0) node[below=of e] {$t$};
\node (D) at (d){};
\node (E) at (e){};
\node (rect) at (2,0.5) [draw,thick,minimum width=2cm,minimum height=1cm] {$\WCET_{i}$};
\node (rect) at (6.5,0.5) [draw,thick,minimum width=2cm,minimum height=1cm] {$\WCET_{i}$};
\node (rect) at (11,0.5) [draw,thick,minimum width=2cm,minimum height=1cm] {$\WCET_{i}$};

\node (rect) at (3.75,0.5) [draw,thick,minimum width=1.5cm,minimum height=1cm] {$\WCET_{j}$};
\node (rect) at (12.75,0.5) [draw,thick,minimum width=1.5cm,minimum height=1cm] {$\WCET_{j}$};

\draw[thick, -Triangle] (0,0) -- (15.5,0) node[font=\scriptsize,below left=3pt and -8pt]{};
\draw (0 cm,3pt) -- (0 cm,-3pt);
\draw[dashed] (0 cm,0pt) -- (0 cm,50pt);
\draw[dashed] (1 cm,-45pt) -- (1 cm,90pt);
\draw[dashed] (5.5 cm,-25pt) -- (5.5 cm,70pt);
\draw[dashed] (4.5 cm,0pt) -- (4.5 cm,50pt);
\draw[dashed] (10 cm,-45pt) -- (10cm,90pt);

\draw[dashed] (14.5 cm,-25pt) -- (14.5 cm,70pt);

\draw[dashed] (9 cm,0pt) -- (9 cm,50pt);

\draw[dashed] (13.5 cm, 0pt) -- (13.5 cm,50pt);

 \draw[arrows={angle 60-angle 60}] (0,1.5) -- node[above]{$\OT_{i}$} (1,1.5) ;
 
 \draw[arrows={angle 60-angle 60}] (1,1.5) -- node[above]{$\DT_{i}$} (4.5,1.5) ;
 \draw[arrows={angle 60-angle 60}] (5.5,1.5) -- node[above]{$\DT_{i}$} (9, 1.5) ;
  \draw[arrows={angle 60-angle 60}] (10,1.5) -- node[above]{$\DT_{i}$} (13.5,1.5) ;
  
 \draw[arrows={angle 60-angle 60}] (1,-0.7) -- node[above]{$\PP_{i}$} (5.5,-0.7);
 \draw[arrows={angle 60-angle 60}] (5.5,-0.7) -- node[above]{$\PP_{i}$} (10,-0.7);
  \draw[arrows={angle 60-angle 60}] (10,-0.7) -- node[above]{$\PP_{i}$} (14.5,-0.7);

  \draw[arrows={angle 60-angle 60}] (1,-1.5) -- node[above]{$\PP_{j}$} (10,-1.5);

 \draw[arrows={angle 60-angle 60}] (1,2.3) -- node[above]{$\PT_{i}$} (5.5,2.3) ;
 \draw[arrows={angle 60-angle 60}] (5.5,2.3) -- node[above]{$\PT_{i}$} (10,2.3) ;
 \draw[arrows={angle 60-angle 60}] (10,2.3) -- node[above]{$\PT_{i}$} (14.5,2.3);
  
   \draw[arrows={angle 60-angle 60}] (1,3.1) -- node[above]{$\MAF_{i}$}(10,3.1);
\end{tikzpicture}
}
\end{center}
  \caption{Real-time characteristics of the system}
  \label{figure:Real-time-characteristics-of-system}
\end{figure}
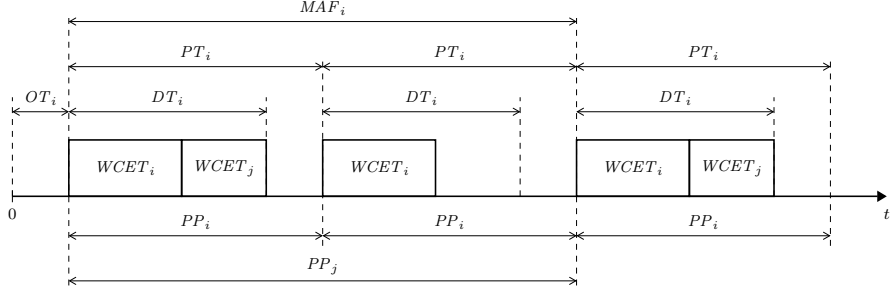
\begin{example}\label{example:RTS}
	Let us illustrate these definitions using \cref{figure:Real-time-characteristics-of-system}.
	 A single thread $\thread_i$ is considered, with offset $\OT_i$, MAF %
	 $\MAF_{i}$, %
	 period $\PT_i$ and deadline %
			$\DT_{i}$.
	 This thread has two processings $\processing_i$
		and $\processing_j$, where $\processing_i$ is characterized by a  WCET $\WCET_i$ and a period $\PP_i$ and $\processing_j$ has WCET $\WCET_j$ and period~$\PP_j$.
\end{example}
A reactivity is of the form $\reactivity_{i} = ((\processing_{i_1} \rightarrow \processing_{i_2} \rightarrow \dots \rightarrow \processing_{i_k}),\DR_{i} )$ where
$\processing_{i_1}, \processing_{i_2} ,  \dots , \processing_{i_k}$ are $k$ processings of $\Processings$,
$(\processing_{i_1} \rightarrow \processing_{i_2} \rightarrow \dots \rightarrow \processing_{i_k})$ denotes a precedence constraint,
and $\DR_{i}$ is the maximum \emph{time of reactivity} for $\reactivity_{i}$: the end of the thread period containing the last processing $\processing_{i_k}$ of the precedence sequence has to be completed before the deadline $\DR_{i}$.
(If a reactivity is satisfied, its precedence constraint is obviously satisfied too.)

\begin{definition}\label{definition:schedulability}
	A system $\RTS$ is \textit{schedulable} if
	\begin{ienumerate}
		\item $\forall$ $\thread_{i}$ $\in$ $\Threads$, the end of each instance of $\thread_{i}$ occurs before its relative deadline $\DT_{i}$.
		\item $\forall$ $\reactivity_{i}$ $\in$ $\Reactivities$, the end of each instance of the thread containing the last processing %
			$\processing_{i_k}$
				of $\reactivity_{i}$ occurs before~$\DR_{i}$.
	\end{ienumerate}
\end{definition}
\ifdefined\LaTeXdiff
\else
\ArXivVersion{
\subsection[Formalization of the case study]{Formalization of the case study}

We formalize in the following the system, with the values given in \cref{figure:example_control_system,figure:example_WCET} and the assignments onto threads given in \cref{figure:typical_solution}.

\subsubsection[Processings]{Processings}

Let $\Processings$ denote the set of processings.
This set can be defined as $\Processings = \{ \procNavi, \procCont, \procMoni, \procGuid \}$, where:

\begin{description}
	\item[Control:] $\procCont = (\WCETCont , \PPCont) = (3, 10) $.
	\item[Guidance:] $\procGuid = (\WCETGuid , \PPGuid) = (15, 60) $.
	\item[Monitoring:] $\procMoni = (\WCETMoni , \PPMoni) = (5, 20) $.
	\item[Navigation:] $\procNavi = (\WCETNavi , \PPNavi) = (1, 5) $.
\end{description}

\subsubsection[Threads]{Threads}

Let $\Threads = \{ \thread_1, \thread_2, \thread_3 \}$ denote the set of threads, with:

\begin{itemize}
\item $\thread_{1} = (\PT_{1} , \OT_{1} , \DT_{1}, \MAF_{1}, \Processings_{1}) = (5, \OT_{1}, \DT_{1}, 10, \{\procNavi, \procCont\}) $.

\item $\thread_{2} = (\PT_{2} , \OT_{2} , \DT_{2}, \MAF_{2}, \Processings_{2}) = (20, \OT_{2}, \DT_{2}, 20, \{\procMoni\})$.

\item $\thread_{3} = (\PT_{3} , \OT_{3} , \DT_{3}, \MAF_{3}, \Processings_{3}) = (60, \OT_{3}, \DT_{3}, 60, \{\procGuid\})$.

\end{itemize}

\subsubsection[Reactivities]{Reactivities}\label{subsubsection:reactivities}

Let $\Reactivities = \{ \reactivity_1, \reactivity_2, \reactivity_3 \}$ denote the set of reactivities, with:

\begin{itemize}
\item $\reactivity_{1} = \big( ( \procNavi \rightarrow \procGuid \rightarrow \procCont),\DR_{1}\big)$ %
	with $\DR_{1} = 150$.
\item $\reactivity_{2} = \big( (\procNavi \rightarrow \procCont) ,\DR_{2}\big)$ %
	with $\DR_{2} = 15$.
\item $\reactivity_{3} = \big( ( \procNavi \rightarrow \procMoni) ,\DR_{3}\big)$ %
	with $\DR_{3} = 55$.
\end{itemize}

}
\fi

\subsection{Objectives}

\begin{figure}[tb]
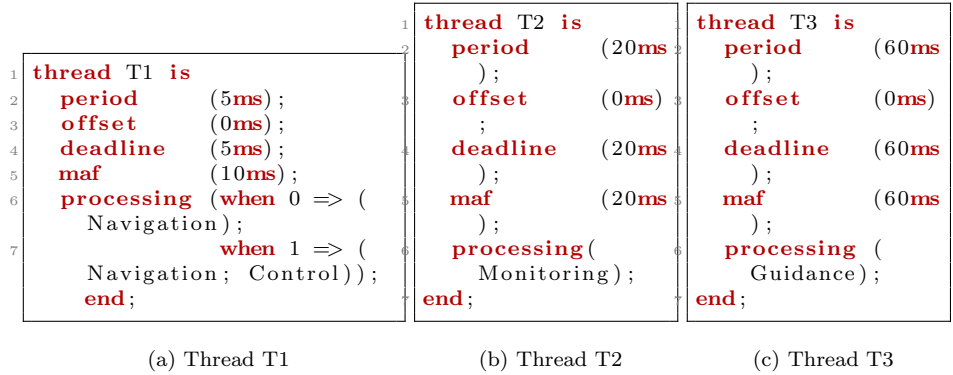

	\subfloat[Thread T1]{
		\begin{minipage}[b]{0.4\textwidth}
\begin{lstlisting}[style=ariane]
thread T1 is
  period     (5ms);
  offset     (0ms);
  deadline   (5ms);
  maf        (10ms);
  processing (when 0 => (Navigation);
              when 1 => (Navigation; Control));end;
\end{lstlisting}
		\end{minipage}
		\label{fig:T1}
	}
	\subfloat[Thread T2]{
		\begin{minipage}[b]{0.27\textwidth}
\begin{lstlisting}[style=ariane]
thread T2 is
  period     (20ms);
  offset     (0ms);
  deadline   (20ms);
  maf        (20ms);
  processing(Monitoring);
end;
\end{lstlisting}
		\end{minipage}
		\label{fig:T2}
	}
	\subfloat[Thread T3]{
		\begin{minipage}[b]{0.27\textwidth}
\begin{lstlisting}[style=ariane]
thread T3 is
  period     (60ms);
  offset     (0ms);
  deadline   (60ms);
  maf        (60ms);
  processing (Guidance);
end;
\end{lstlisting}
		\end{minipage}
		\label{fig:T3}
	}
		\caption{A typical solution of the flight control scheduling problem}
 		\label{figure:typical_solution}
\end{figure}

\ArXivVersion{%
	Let us summarize the problems we address in this paper.
}
Our problems take as input a real-time system, i.e.:
\begin{ienumerate}
	\item a list of processings with their WCET (for example \cref{figure:example_WCET}) and period, and their input or output data (for example \cref{figure:example_control_system});
	\item a set of reactivities (for example \cref{figure:reactivities});
	\item an allocation of processings on threads, with period, offset, deadline and MAF for each thread
		(for example \cref{figure:typical_solution}).
\end{ienumerate}

\begin{remark}
	Observe in \cref{figure:typical_solution}
	that the harmonic assumption on threads is respected, with threads ordered by increasing frequency as follows: T3, T2, T1.
\end{remark}

The first problem is to formally \emph{verify} the schedulability of the real-time system:

\smallskip

\defProblem
	{Scheduling verification}
	{a real-time system}
	{formally verify that $\RTS$ is schedulable.}

\medskip

Recall that schedulability also ensures that all %
	reactivity constraints are met (from \cref{definition:schedulability}).

The second problem assumes that some constants of the real-time system (deadlines, periods, offsets, WCET…)\ become unknown.
The real-time system can then be seen as a \emph{partially specified} or \emph{abstract} system.

In this work, we assume that the offsets and deadlines of each thread are unknown; that is, some of the values in \cref{figure:typical_solution}
are not known anymore.
The scheduling synthesis problem for our flight control system consists thus in computing the offsets and deadlines of each thread in order to fulfill the required reactivities.

\smallskip

\defProblem
	{Scheduling synthesis}
	{a real-time system, a set of unknown constants}
	{exhibit valuations for the unknown constants such as $\RTS$ is schedulable. %
	}

Recall that our synthesis problem still considers as input the periods; therefore offsets and deadlines are the main results of interest.

\section{Parametric stopwatch automata}\label{section:PSA}

\parORsubsection{Clocks, parameters, constraints}
We assume a set~$\Clock = \{ \clock_1, \dots, \clock_\ClockCard \} $ of \emph{clocks}, \ie{} real-valued variables that evolve at the same rate.
A clock valuation is a function $\clockval : \Clock \rightarrow \Rgeqzero$. 
We write $\ClocksZero$ for the clock valuation assigning $0$ to all clocks.
Given $\resets \subseteq \Clock$, we define the \emph{reset} of a valuation~$\clockval$, denoted by $\reset{\clockval}{\resets}$, as follows:
$\reset{\clockval}{\resets}(\clock) = 0$ if $\clock \in \resets$, and
$\reset{\clockval}{\resets}(\clock)=\clockval(\clock)$ otherwise.
Given a valuation~$\clockval$, $d \in \grandrplus$ and $\Clock' \subseteq \Clock$, we define the \emph{time-elapsing of~$\clockval$ by~$d$ except for clocks in~$\Clock'$}, denoted by $\timelapseStop{\clockval}{d}{\Clock'}$, as the clock valuation such that
\ifdefined\VersionArXiv
\[\timelapseStop{\clockval}{d}{\Clock'}(\clock) = \begin{cases} 
      \clockval(\clock) & \text{if } \clock \in \Clock' \\
      \clockval(\clock) + d & \text{otherwise}\\
   \end{cases}
\]
\else
	\(\timelapseStop{\clockval}{d}{\Clock'}(\clock) =\clockval(\clock) \) if \(\clock \in \Clock' \) and \(\clockval(\clock) + d \) otherwise.
\fi

We assume a set~$\Param = \{ \param_1, \dots, \param_\ParamCard \} $ of
\emph{parameters}, \ie{} unknown constants.
A parameter {\em valuation} $\pval$ is a function $\pval : \Param \rightarrow
\grandqplus$.
We denote ${\compOp} \in \{<, \leq, =, \geq, >\}$.
A guard~$\guard$ is a
constraint over $\Clock \cup \Param$ defined by a conjunction of inequalities
of the form $\clock \compOp d$ or $\clock \compOp \param$, with
$\clock\in\Clock$, $d \in \grandn$ and $\param \in \Param$.
Given a guard
$\guard$, we write~$\clockval\models\pval(\guard)$ if the expression obtained by replacing in~$\guard$ each~$\clock\in\Clock$ by~$\clockval(\clock)$ and each~$\param\in\Param$ by~$\pval(\param)$ evaluates to true.

\parORsubsection{Parametric stopwatch automata}
Parametric timed automata (PTA) extend timed automata with parameters within guards and invariants in place of integer constants~\cite{AHV93}.
For many real-time systems, especially when they are subject to preemptive scheduling, PTA are not sufficiently expressive.
As a result, we will use here an extension of PTA with stopwatches~\cite{CL00}, namely parametric stopwatch automata~\cite{SSLAF13}.

\begin{definition}[PSA]\label{def:PSA}%
	A \emph{parametric stopwatch automaton} (PSA) $\A$ is a tuple \mbox{$\A = (\Actions, \Loc, \locinit, \Clock, \Param, \invariant, \stopFunction, \Edges)$}, where:
    \begin{ienumerate}
		\item $\Actions$ is a finite set of actions,
		\item $\Loc$ is a finite set of locations,
		\item $\locinit \in \Loc$ is the initial location,
		\item $\Clock$ is a finite set of clocks,
		\item $\Param$ is a finite set of parameters,
        \item $\invariant$ is the invariant, assigning to every $\loc\in \Loc$ a guard $\invariant(\loc)$,
        \item $\stopFunction$ is the stop function $\stopFunction : \loc \rightarrow 2^\Clock$, assigning to every $\loc \in \Loc$ a set of stopped clocks,
		\item $\Edges$ is a finite set of edges  $\edge = (\loc,\guard,\action,\resets,\loc')$
		where~$\loc,\loc'\in \Loc$ are the source and target locations,
			$\guard$ is a guard,
			$\action \in \Actions$,
			and
			$\resets\subseteq \Clock$ is the set of clocks to be reset.
    \end{ienumerate}
\end{definition}

Stopwatch automata can be composed as usual using parallel composition on synchronized actions.
Note that our clocks are \emph{shared} by default, \ie{} a same clock (\ie{} with the same name) can be read, stopped or reset in several automata.
The same applies to parameters.

Given a parameter valuation~$\pval$ and PSA~$\A$, we denote by $\valuate{\A}{\pval}$ the non-parametric structure where, for each parameter~$\param\in\Param$, all occurrences of~$\param$ have been replaced by~$\pval(\param)$.
Any structure $\valuate{\A}{\pval}$ is also a \emph{stopwatch automaton}~\cite{CL00}.
If $\stopFunction(\loc) = \emptyset$ for all $\loc \in \Loc$, then by assuming a rescaling of the constants (multiplying all constants in $\valuate{\A}{\pval}$ by the least common multiple of their denominators), we obtain an equivalent (integer-valued) TA, as defined in~\cite{AD94}.

Let us now recall the concrete semantics of stopwatch automata.

\begin{definition}%
	Given a PSA $\A = (\Actions, \Loc, \locinit, \Clock, \Param, \invariant, \stopFunction, \Edges)$,
	and a parameter valuation~\(\pval\),
	the semantics of $\valuate{\A}{\pval}$ is given by the timed transition system (TTS) $(\States, \sinit, \flecheRel)$, with
	\begin{oneenumerate}
		\item $\States = \{ (\loc, \clockval) \in \Loc \times \Rgeqzero^\ClockCard \mid \clockval \models \valuate{\invariant(\loc)}{\pval} \}$, %
		\item $\sinit = (\locinit, \ClocksZero) $,
		\item $\flecheRel$ consists of the discrete and (continuous) delay transition relations:
		\begin{ienumerate}
			\item discrete transitions: $(\loc,\clockval) \longueflecheRel{\edge} (\loc',\clockval')$, %
				if $(\loc, \clockval) , (\loc',\clockval') \in \States$, and there exists $\edge = (\loc,\guard,\action,\resets,\loc') \in \Edges$, such that $\clockval'= \reset{\clockval}{\resets}$, and $\clockval\models\pval(\guard$).
			\item delay transitions: $(\loc,\clockval) \longueflecheRel{d} (\loc, \timelapseStop{\clockval}{d}{\stopFunction(\loc)} )$, with $d \in \Rgeqzero$, if $\forall d' \in [0, d], (\loc, \timelapseStop{\clockval}{d'}{\stopFunction(\loc)}) \in \States$.
		\end{ienumerate}
	\end{oneenumerate}
\end{definition}

\section{Specifying the system}\label{section:specification}

Since the seminal work of Liu and Layland in~\cite{LL73}, an abundant number of methods and tools have been designed to check the schedulability of real-time systems.
However, while some aspects are reasonably easy (\FPS{}, no mixed-criticality), the problem we address here is not typical for several reasons:
\begin{ienumerate}
	\item offsets may be non-null;
	\item the executed processings may differ depending on the cycle;
	\item the reactivities must always be met, and therefore define new, non-classical timing constraints; and, perhaps most importantly,
	\item the admissible values for deadlines and offsets may not be known. Only the global end-to-end reactivity is specified.
\end{ienumerate}

As a consequence, we choose to follow a \emph{model checking} based method.
Model checking is known for being more expressive than analytical methods, at the cost of performance or even decidability.
We show here that, although we use an undecidable formalism, we do get exact results for the instance of the problem we consider.
We indeed rely on a procedure (``reachability synthesis'', formalized in \eg{} \cite{JLR15}) which is not guaranteed to terminate---but is correct whenever it does.\label{newtext:undecformalism}

We present in the remainder of this section our modeling of the verification and the synthesis problem using PSA.
This formalism has several advantages.
First, it is helpful to model concurrent aspects of the system (different threads and processings running concurrently).
Second, stopwatches can be used to model preemption.
Third, parameters can be used to model the unknown constants, and solve the synthesis problem.

For now, we consider that there are no context switches in the system.
We will discuss in \cref{section:switch} how to introduce them.

\subsection{Architecture of the solution}

\subsubsection{A modular solution}
To model the system, we use the concurrent structure of parametric stopwatch automata so as to build a modular solution: that is, each element (thread, processing, scheduling policy) and each constraint (reactivity) is defined by a dedicated PSA.
These automata are then composed by usual parallel composition on synchronization actions.

This makes our solution modular in the sense that, in the case of a modification in the system (\eg{} the scheduling policy), we can safely replace one PSA with another (\eg{} the \FPS{} scheduler automaton with another scheduler PSA) without impacting the rest of the system.

\subsubsection{Encoding elements and constraints as automata}
We will model each processing activation as a PSA.
These automata ensure that processings are activated periodically with their respective period and initial offset.

In addition, we will create one PSA for each thread: the purpose of these automata is to %
	ensure that the processings associated with each thread are executed at the right time.
In the case of our concrete problem, we assign both the Navigation and Control processings to thread~T1, the monitoring process to~T2 and the guidance processing to~T3.

The reactivities also follow the concept of modularity.
That is, each reactivity is \emph{tested} using a single PSA.
By testing (as in~\cite{ABBL03}), we mean that a reactivity fails iff a special location is reached.
Therefore, ensuring the validity of the reactivities is equivalent to the unreachability of these special locations.

Finally, we will specify a scheduler automaton that encodes the scheduling policy between the different threads (in our problem, recall that the scheduling policy is fixed priority scheduling (\FPS{})).

We give more details on each of these automata in the following.

\subsection{Modeling periodic processing activations}
To model the periodicity of the processings, we create one PSA for each processing activation.
This PSA simply performs the activations in a periodic manner.
Activations are modeled by a synchronization action that is used to communicate with other automata (typically the thread automaton).
For example, the activation of the Control processing is denoted by \styleact{actControl}; this action will be used to synchronize between the Control activation automaton with other automata (\eg{} the threads\ArXivVersion{ or the scheduler}).

In addition, the period processing activation automaton detects whether a processing missed its\ArXivVersion{ (implicit)} deadline (equal to its period);
	that is, we assume that a processing that has not finished by its next period is a situation corresponding to a deadline miss.
\ifdefined\LaTeXdiff\else\ArXivVersion{

}\fi%
Each %
	automaton features a single clock.

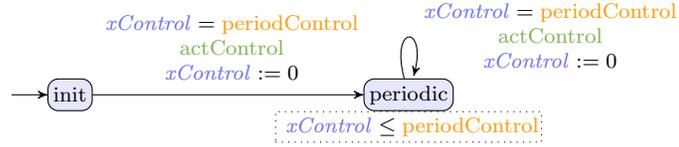
\begin{figure}
	\centering
	\scriptsize
	\begin{tikzpicture}[scale=1.5, auto, ->, >=stealth']
 
		\node[location, initial] at (0,0) (l1) {\styleloc{init}};

		\node[location] at (3, 0) (l2) {\styleloc{periodic}};
		\node [invariant,below] at (l2.south) {\begin{tabular}{@{} c @{\ } c@{} }& $ \styleclock{xControl} \leq \styleparam{periodControl}$\\\end{tabular}};

		\path (l1) edge node[above]{\begin{tabular}{@{} c @{\ } c@{} }
		& $ \styleclock{xControl} = \styleparam{periodControl}$\\
		 & $\styleact{actControl}$\\
		 & $\styleclock{xControl}:=0$\\
		\end{tabular}} (l2);

		\path (l2) edge[loop above] node[right]{\begin{tabular}{@{} c @{\ } c@{} }
		& $ \styleclock{xControl} = \styleparam{periodControl}$\\
		 & $\styleact{actControl}$\\
		 & $\styleclock{xControl}:=0$\\
		\end{tabular}} (l2);
	\end{tikzpicture}
	
	\caption{Automaton \stylePTA{periodicControl}}
	\label{figure:periodicControl}
\end{figure}

We present in \cref{figure:periodicControl} a simplified version of the \stylePTA{periodicControl} automaton, modeling the periodic activation of the Control processing.\footnote{%
	Among the simplifications, we do not represent the check for the deadline miss.
}
This automaton uses one clock \styleclock{xControl} and one parameter \styleparam{periodControl}.
The clock \styleclock{xControl} is used to measure the time between any two consecutive processing activations; it is never stopped.
Note that the period \styleparam{periodControl} is known beforehand, and is therefore not strictly speaking a parameter, but that makes our solution both more generic and more readable (in \imitator{}, a parameter can be statically instantiated to a constant\ArXivVersion{ before running the analysis}).

The initial location is \styleloc{init}: from then, the first occurrence of Control is immediately activated (action \styleact{actControl}), and the automaton enters the \styleloc{periodic} location.
Then, exactly every \styleparam{periodControl} time units (guard $\styleclock{xControl} = \styleparam{periodControl}$), another instance of Control is activated.

\subsection{Modeling threads}

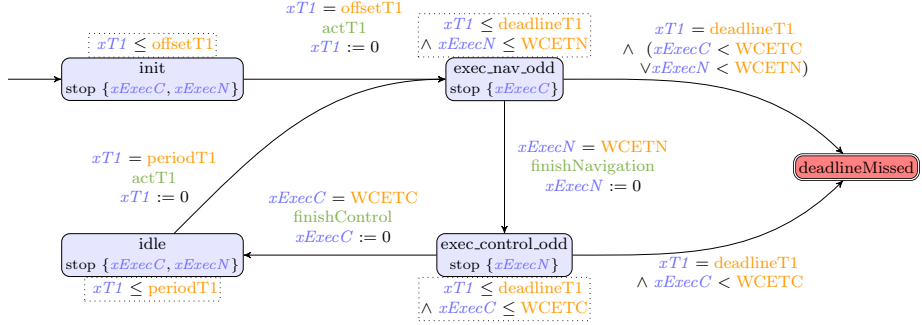
\begin{figure*}
	\centering
	\ifdefined\VersionArXiv
		\newcommand{\scalefactorTone}{.78}
	\else
		\newcommand{\scalefactorTone}{.9}
	\fi
	\scalebox{\scalefactorTone}{
	\begin{tikzpicture}[scale=2, auto, ->, >=stealth']
 
		\node[location, initial] at (0,0) (init) [align=center] {\styleloc{init}\\{\scriptsize stop $\{ \styleclock{xExecC} , \styleclock{xExecN} \}$}};
		\node [invariant, above] at (init.north) {\begin{tabular}{@{} c @{\ } c@{} }& $ \styleclock{xT1} \leq \styleparam{offsetT1}$\\\end{tabular}};
 
		\node[location] at (3, 0) (execNav) [align=center] {\styleloc{exec\_nav\_odd}\\{\scriptsize stop $\{ \styleclock{xExecC} \}$}};
		\node [invariant,above] at (execNav.north) {\begin{tabular}{@{} c @{\ } c@{} }& $ \styleclock{xT1} \leq \deadlineTone{}$ \\ $\land$ & $ \styleclock{xExecN} \leq \styleparam{WCETN}$ \end{tabular}};
 
		\node[location] at (3, -1.5) (execCon) [align=center] {\styleloc{exec\_control\_odd}\\{\scriptsize stop $\{ \styleclock{xExecN} \}$}};
		\node [invariant, below] at (execCon.south) {\begin{tabular}{@{} c @{\ } c@{} }& $ \styleclock{xT1} \leq \deadlineTone{}$ \\ $\land$& $ \styleclock{xExecC} \leq \styleparam{WCETC}$\end{tabular}};
 
		\node[location] at (0, -1.5) (idle) [align=center] {\styleloc{idle}\\{\scriptsize stop $\{ \styleclock{xExecC} , \styleclock{xExecN} \}$}};
		\node [invariant,below] at (idle.south) {\begin{tabular}{@{} c @{\ } c@{} }& $ \styleclock{xT1} \leq \styleparam{periodT1}$ \end{tabular}};
 
		\node[location, bad] at (6, -0.75) (missed) [align=center] {\styleloc{deadlineMissed}};

		\path (init) edge node[above,align=center]{
		$\styleclock{xT1} = \styleparam{offsetT1}$\\
		$\styleact{actT1}$\\
		$\styleclock{xT1}:=0$\\
		} (execNav);
 
		\path (execNav) edge node[right]{\begin{tabular}{@{} c @{\ } c@{} }
		& $ \styleclock{xExecN} = \styleparam{WCETN}$\\
		 & $\styleact{finishNavigation}$\\
		 & $\styleclock{xExecN}:=0$\\
		\end{tabular}} (execCon);
		
		\path (execCon) edge node[above]{\begin{tabular}{@{} c @{\ } c@{} }
		& $ \styleclock{xExecC} = \styleparam{WCETC}$\\
		 & $\styleact{finishControl}$\\
		 & $\styleclock{xExecC}:=0$\\
		\end{tabular}} (idle);

		\path (idle) edge[out=45, in=180] node[below left, xshift=-35, yshift=-10, align=center]{
		$ \styleclock{xT1} = \styleparam{periodT1}$\\
		$\styleact{actT1}$\\
		$\styleclock{xT1}:=0$\\
		} (execNav);
		
		\path (execNav) edge[out=0, in=135] node[above]{\begin{tabular}{@{} c @{\ } c@{} }
		& $ \styleclock{xT1} = \styleparam{deadlineT1}$\\
 		 $\land$ & $ (\styleclock{xExecC} < \styleparam{WCETC}$\\
	 		& $ \lor \styleclock{xExecN} < \styleparam{WCETN})$\\
		\end{tabular}} (missed);
		
		\path (execCon) edge[out=0, in=-135] node[below]{\begin{tabular}{@{} c @{\ } c@{} }
		& $ \styleclock{xT1} = \styleparam{deadlineT1}$\\
 		 $\land$ & $ \styleclock{xExecC} < \styleparam{WCETC}$\\
		\end{tabular}} (missed);
	\end{tikzpicture}
}
	
	\caption{Fragment of automaton \stylePTA{threadT1}}
	\label{figure:threadt1}
\end{figure*}

We create one PSA for each thread.
Each of these automata contains one clock for the thread (used to measure the thread period and offset), as well as one clock per processings assigned to the thread.
These processings clocks are used to measure the amount of time spent on executing these processings; these clocks can be stopped (they are therefore \emph{stopwatches}, strictly speaking) when the processor was preempted for a higher priority task.
For example in \cref{figure:threadt1}, the thread automaton \stylePTA{threadT1} contains \styleclock{xT1} (the thread clock), as well as \styleclock{xExecControl} and \styleclock{xExecNavigation} (the clocks associated to the processings of~T1).
Parameters include the offset, period, and deadline of the thread, but also the WCETs of the processings assigned to this thread.

The thread automaton is responsible for:
\begin{oneenumerate}
	\item encoding the initial thread offset, \ie{} starting the periodic thread activation only after the offset;
	\item performing the periodic thread activation;
	\item executing the processings associated with the thread; and
	\item detecting the deadline misses.
\end{oneenumerate}

The clocks associated with the processings are used to measure the execution time of these processings: they are in fact stopped most of the time, except when the thread is actively executing the processing.
This is in contrast with the clocks associated with the processing activation automaton, which are never stopped, as they measure a period.
Then, a deadline miss occurs if the clock measuring the thread period reaches the deadline (recall that the deadline is less than or equal to the period, and therefore we can use the same clock), while the clock measuring a processing execution time is strictly less than its WCET.

We give in \cref{figure:threadt1} a fragment of the automaton \stylePTA{threadT1}.
We only give the odd cycle, as this is the most interesting; that is, we removed the fragment corresponding to the even cycle (only executing Navigation) between locations \styleloc{init} and \styleloc{exec\_nav\_odd}\ArXivVersion{ (and the transition from \styleloc{idle} should go to the removed \styleloc{exec\_nav\_even} location)}.
The automaton uses several synchronization variables, notably the end of the processings (\eg{} \styleact{finishControl}), but also the start and end of the concerned thread (\eg{} \styleact{actT1} and \styleact{endT1}, not depicted in the simplified version in \cref{figure:threadt1}).
We also abbreviate some variable names to save space (\eg{} $\styleclock{xExecC}$ for $\styleclock{xExecControl}$  and  \styleclock{xExecN} for \styleclock{xExecNavigation} or \styleparam{WCETN} for \styleparam{WCETNavigation}).

First, the automaton waits for the offset: that is, it stays in \styleloc{init} exactly \styleparam{offsetT1} time units.
Then, it executes the first processing of the odd cycle, \ie{} Navigation: it stays in \styleloc{exec\_nav\_odd} until completion, \ie{} for \styleparam{WCETNavigation} time units.\footnote{%
	In the full model, we can allow for a best case execution time, in which case the duration is nondeterministically chosen in the interval $[\styleparam{BCETNavigation}, \styleparam{WCETNavigation}]$.
}
Note that this is the only location where \styleclock{xExecNavigation} is elapsing, \ie{} is not stopped, as it measures the execution time.
Then, upon completion of the Navigation processing, the automaton moves to \styleloc{exec\_control\_odd}, where Control is executed.
Upon completion, it moves to \styleloc{idle}, and waits until the clock $\styleclock{xT1}$ reaches its period.
Then, the cycle restarts and so on.

In addition, at any time, possible deadline misses are checked for.
A deadline miss occurs on an odd cycle while execution Navigation whenever $\styleclock{xT1} = \deadlineTone$ and either $\styleclock{xExecControl} < \styleparam{WCETControl}$ or $\styleclock{xExecNavigation} < \styleparam{WCETNavigation}$.\footnote{%
	This encoding is not necessarily optimal.
	In fact, on odd cycles, as Navigation is executed first, and followed by Control, a deadline miss can be detected earlier, \ie{} if Navigation is still executed, but there is not enough time to finish the execution of Navigation and that of Control: that is, an optimized deadline miss condition could be $\styleclock{xT1} + \styleparam{WCETControl} = \deadlineTone$ and $\styleclock{xExecNavigation} < \styleparam{WCETNavigation}$.
	This optimization has not been implemented, so as to leave the model (relatively) simple and maintainable, but could be tested in the future.
}
When executing Control, only the execution time of Control needs to be checked.

\ArXivVersion{%
\begin{remark}
Our model is in fact more complicated as, for sake of modularity, we make no assumption in the thread automaton on how the other automata behave, notably the processings activation automata.
	Therefore, we allow for processings to be activated at any time, which must be taken care of in the thread automaton.
\end{remark}
}

\subsection{Modeling the \FPS{} scheduler}

The \FPS{} scheduler is modeled using an additional PSA.
It reuses existing works from the literature (\eg{} \cite{FKPY07,SSLAF13}), and does not represent a significant original contribution.
We mainly reuse the scheduler encoding of \cite{SSLAF13}, which consists of an automaton synchronizing with the rest of the system on the start and end task synchronization actions as well as the task activation actions.
Whenever a new task is activated, the scheduler decides what to do depending on its current state and the respective priorities of the new and the executing tasks (if any).

Nevertheless, we had to modify this encoding due to the fact that existing scheduler automata simply schedule tasks: in our setting, the scheduler schedules both the threads and the threads' processings.
Among the various modifications, in case of preemption, our scheduler does not stop the clocks measuring the execution times of the preempted threads (because such clocks do not exist), but stop the clocks measuring the execution times of the \emph{processings %
	deployed on %
	the preempted threads}.\label{newtext:deployed}

\begin{figure}
	\centering
		\begin{tikzpicture}[scale=.8, xscale=2.5, yscale=1, auto, ->, >=stealth']
 
		\node[location, initial] at (0, 1) (idle) [align=center] {\styleloc{idle}  }; %
		\node [invariant, above left,align=center] at (idle.west) {};
 
		\node[location] at (2,2) (execT1) [align=center] {\styleloc{execT1}};

		\node[location] at (4, 1) (execT1waitT2) [align=center] {\styleloc{execT1waitT2} \\ stop$\{ \styleclock{xexecM} \}$};

		\node[location] at (2, 0) (execT2) [align=center] {\styleloc{execT2}};

		\path (idle) edge[bend angle=20,bend right] node[above,yshift=-2]{\begin{tabular}{@{} c @{\ } c@{} }
		 & $\styleact{actT1}$\\
		\end{tabular}} (execT1);

		\path (idle) edge[bend angle=20,bend left] node[below, yshift=3]{\begin{tabular}{@{} c @{\ } c@{} }
		 & $\styleact{actT2}$\\
		\end{tabular}} (execT2);

		\path (execT1) edge[bend right] node[above left]{\begin{tabular}{@{} c @{\ } c@{} }
		 & $\styleact{endT1}$\\
		\end{tabular}} (idle);

		\path (execT1) edge node{\begin{tabular}{@{} c @{\ } c@{} }
		 & $\styleact{actT2}$\\
		\end{tabular}} (execT1waitT2);

		\path (execT1waitT2) edge[bend left] node[]{\begin{tabular}{@{} c @{\ } c@{} }
		 & $\styleact{endT1}$\\
		\end{tabular}} (execT2);

		\path (execT2) edge[bend left] node{\begin{tabular}{@{} c @{\ } c@{} }
		 & $\styleact{endT2}$\\
		\end{tabular}} (idle);

		\path (execT2) edge[bend left] node[above left, yshift=-6]{\begin{tabular}{@{} c @{\ } c@{} }
		 & $\styleact{actT1}$\\
		\end{tabular}} (execT1waitT2);

	\end{tikzpicture}
	
	\caption{Encoding the \FPS{} scheduler (simplified version)}
	\label{figure:scheduler}
\end{figure}
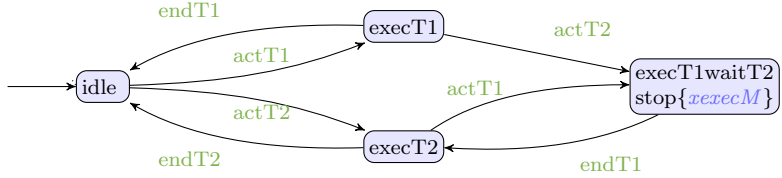

We give in \cref{figure:scheduler} an example of such a scheduler in a simplified version, with only two threads T1 and~T2\ArXivVersion{; the full scheduler is of course more complete}.
If any of the two threads get activated (\styleact{actT1} or \styleact{actT2}), the scheduler starts executing them.
If a second thread gets activated, the highest priority thread (T1) is executed, while T2 is put on the waiting list (which is encoded in location \styleloc{execT1waitT2}).
This is the location responsible for stopping the clock of the (only) processing of~T2, \ie{} Monitoring (clock~$\styleclock{xexecM}$).
Only after T1 has completed (\styleact{endT1}), T2 can execute.
Our real scheduler is in fact significantly more complex as it has to cope with three threads, but also with special cases\ArXivVersion{ such as the activation of a new thread activation of~$\thread_i$ while executing a previous instance of~$\thread_i$, etc}.

\subsection{Reachability synthesis}\label{ss:reachability}

Finally, the system is schedulable if none of the ``bad'' locations (corresponding to deadline misses, \eg{} in the thread automata) is reachable.
If all parameters are valuated, the system is a TA, and schedulability reduces to reachability checking.
If some parameters are free (\ie{} the analysis is parametric), the set of valuations for which the system is schedulable exactly corresponds to the valuations for which these bad locations are unreachable, \ie{} the complement of the valuations set result of reachability synthesis.
This guarantees our method correctness.

\section{Compositional verification of reactivities}\label{section:compositional}

An originality of our work---which among other reasons, notably the timing parameters, justifies our choice to use model checking---is the encoding of \emph{reactivities}.
Indeed, our goal is to verify a system, or synthesize valuations, for which all reactivities are met.
\ifdefined\LaTeXdiff\else\ArXivVersion{

}\fi%
How to properly encode reactivities turned out rather subtle.
Let us first exemplify the complexity of the definition of reactivities.

\begin{example}\label{example:reactivity:NM}
	Consider the third reactivity in \cref{figure:reactivities} (abbreviated by NM in the following) that requires that any data transmission Meas $\rightarrow$ Navigation $\rightarrow$ Monitoring $\rightarrow$ Safeguard must always be less than 55\,ms.
	Recall that data are transmitted upon the end of a \emph{thread} period.

	We can see this reactivity as the start of a timer at the beginning of the last thread period of an execution of Navigation that completed before the end of an execution of task~T1, where T1 is such that it is the last execution of~T1 the period of which ends before the start of an execution of~Monitoring; then, the timer stops following the end of the period of an execution of~T1 immediately following the end of the period of~T3 corresponding to the end of the aforementioned execution of~Monitoring.
	At the end, the timer must be less than~55\,ms.
	
	In other words, this reactivity requires that any following sequence of actions should take less than 55\,ms:
	\styleact{actT1},
	\styleact{startNavigation} followed by \styleact{endNavigation} (without any occurrence of \styleact{startNavigation} in between)
	followed by \styleact{endT1},
	followed by \styleact{actT3} (without any occurrence of \styleact{endT1} in between),
	\styleact{startMonitoring} followed by \styleact{endMonitoring} (without any occurrence of \styleact{startMonitoring} in between),
	followed by \styleact{endT3}.
\end{example}

Encoding reactivities is arguably the most technical part of our solution, and we tried multiple methods (either incorrect or that represented a too large overhead) before converging to this solution.
Nevertheless, the solution we chose still represents a large overhead, as we will see in \cref{section:experiments}.

In our solution, each reactivity is encoded as a sort of \emph{observer} automaton~\cite{ABBL03,Andre13ICECCS}; an observer automaton observes the system behavior without interfering with it.
That is, it can read clocks, and synchronize on synchronization actions, but without impacting the rest of the systems; in particular, it must be non-blocking (except potentially once the property verified by the observer is violated).
In addition, an observer often reduces to \emph{reachability} analysis: the property encoded by the observer is violated iff a special location of the observer is reachable.

Each reactivity automaton uses a single (local) clock used to check the reactivity constraint, and synchronizes with the rest of the system on (some) synchronization labels encoding the start and end of processings and tasks.

In fact, we deviate from the principle of observer automaton by allowing it to block in some cases.
Indeed, a key point in the definition of reactivities in our problem is the communication between threads as exemplified in \cref{example:reactivity:NM}.
In order to allow a generic solution for reactivities, and due to the fact that some timing parameters are unknown, we cannot make assumptions on the respective ordering of processings \wrt{} each other.
Therefore, when a given processing is faster than another one (\eg{} Navigation is faster than Guidance), it is not possible to know \emph{a priori} which instance of the fast processing (\eg{} Navigation) will effectively transmit its data to the following slower processing (\eg{} Monitoring).
As a consequence, our observer will nondeterministically ``guess'' from which instance of the slower processing to start its timer: this is achieved by a nondeterministic choice in the initial location of the automaton.
If the guess is wrong, the observer ``blocks'' the system (impossibility to fire a transition or let time elapse).
Note that, while blocking is usually not an admissible feature of observer automata, this is harmless in this case as, due to the nondeterministic guess and the fact that model checking explores all choices, all possible behaviors of the system are still explored by our solution.

\begin{example}\label{example:reactivity:NM2}
	Consider again reactivity NM from \cref{example:reactivity:NM}.
	Consider a given instance of Navigation.
	If a second full instance of Navigation (including the end of thread T1) is observed before the start of~T2, our observer made a wrong guess, and the observer clock is not measuring a proper reactivity, as the instance of Navigation on which the clock should be started must be the last completed instance before the start of~T2.
	In that case, the observer simply blocks.
\end{example}

\todo{préciser que les ECNA seraient un bon formalisme pour récupérer le dernier appel ?!}

\subsection{Observer construction}

Our solution consists in translating the sequence of starting and ending actions of threads and processings following the definition of the reactivities, while forbidding some actions in some locations to ensure the proper encoding of the definition of thread communication and reactivities.
In addition, a clock measuring the reactivity is started upon the (nondeterministic) activation of the first thread, and is checked against the reactivity nominal maximum time upon completion of the last thread.
If this maximum time constraint is violated, the observer enters a special ``bad'' location.
This observer violation location is added to the list of ``bad'' locations in \cref{ss:reachability} when performing reachability synthesis.

\begin{figure*}
	\centering
	\ifdefined\VersionArXiv
		\newcommand{\scalefactorReactNM}{.74}
	\else
		\newcommand{\scalefactorReactNM}{.85}
	\fi
	\scalebox{\scalefactorReactNM}{
\begin{tikzpicture}[scale=1.3, auto, ->, >=stealth']

	\node[location, initial] at (0,0) (init) [align=center] {\styleloc{init}};

	\node[location] at (2, 0) (execT1) [align=center] {\styleloc{exec\_T1}};

	\node[location] at (4, 0) (execN) [align=center] {\styleloc{exec\_N}};

	\node[location] at (6, 0) (endingT1) [align=center] {\styleloc{ending\_T1}};

	\node[location] at (8, 0) (waitT3) [align=center] {\styleloc{wait\_T3}};

	\node[location] at (10, 0) (execT3) [align=center] {\styleloc{exec\_T3}};

	\node[location] at (10, -1) (execM) [align=center] {\styleloc{exec\_M}};

	\node[location] at (8, -1) (endingT3) [align=center] {\styleloc{ending\_T3}};

	\node[location] at (5, -1) (good) [align=center] {\styleloc{good}};

	\node[location] at (7, -.5) (bad) [align=center] {\styleloc{bad}};

	\path (init) edge node[above]{\styleact{actT1}} node[below]{$\styleclock{xNM} := 0$} (execT1);
	\path (init) edge[loop above] node[above]{\styleact{\Actions_{NM}}} (init);

	\path (execT1) edge node[auto]{\styleact{startN}} (execN);
	\path (execT1) edge[loop above] node[above]{\allActExcept{\styleact{startN}}} (execT1);
	
	\path (execN) edge node[auto]{\styleact{endN}} (endingT1);
	\path (execN) edge[loop above] node[above]{\allActExcept{\styleact{endN}}} (execN);
	
	\path (endingT1) edge node[auto]{\styleact{endT1}} (waitT3);
	\path (endingT1) edge[loop above] node[above]{\allActExcept{\styleact{endT1}}} (endingT1);
	
	\path (waitT3) edge node[auto]{\styleact{actT3}} (execT3);
	\path (waitT3) edge[loop above] node[above]{\allActExcept{\styleact{actT3}, \styleact{endT1}}} (waitT3);
	
	\path (execT3) edge node[auto]{\styleact{startM}} (execM);
	\path (execT3) edge[loop above] node[right]{\allActExcept{\styleact{startM}}} (execT3);
	
	\path (execM) edge node[auto]{\styleact{endM}} (endingT3);
	\path (execM) edge[loop right] node[right]{\allActExcept{\styleact{endM}}} (execM);
	
	\path (endingT3) edge node[auto,align=center]{$\styleclock{xNM} \leq \styleparam{reacNM}$\\\styleact{endT3}} (good);
	\path (endingT3) edge[loop below] node[right]{\allActExcept{\styleact{endT3}}} (endingT3);

	\path (endingT3) edge node[right,yshift=.5em, xshift=-1em,align=center]{$\styleclock{xNM} > \styleparam{reacNM}$\\\styleact{endT3}} (bad);
\end{tikzpicture}
}
	
	\caption{Encoding reactivity Navigation $\rightarrow$ Monitoring}
	\label{figure:reactivityNM}
\end{figure*}
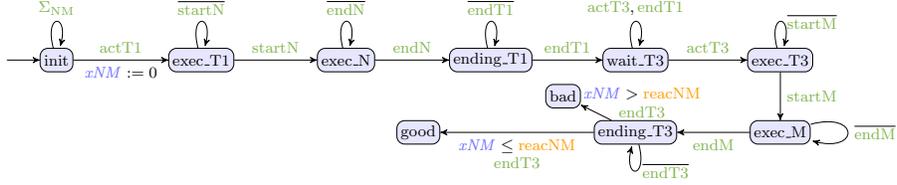
\begin{example}\label{example:reactivity:NM-PTA}
	We give the observer automaton corresponding to reactivity NM in \cref{figure:reactivityNM}.
	We abbreviate\ArXivVersion{ in \cref{figure:reactivityNM}} the names of processings (N and~M stand for Navigation and Monitoring respectively).
	The only clock is \styleclock{xNM} while \styleparam{reacNM} denotes the maximum nominal reactivity for NM (55\,ms here).
	$\Actions_{NM}$ stands for this automaton alphabet; given $\action \in \Actions_{NM}$, $\allActExcept{\action}$ denotes $\Actions_{NM} \setminus \{ \action \}$ (we extend this notation to sets of actions).
	In addition, whenever $\styleclock{xNM} > \styleparam{reacNM}$ occurs in any location (except the initial location), a transition leads to the special ``\styleloc{bad}'' location (these transitions are not depicted in \cref{figure:reactivityNM} for sake of clarity).

	The nondeterministic choice is encoded in the initial location where, upon action~\styleact{actT1}, the automaton can either self-loop in \styleloc{init}, or go to \styleloc{actT1} to try to measure the reactivity from this instance of~T1.
	The blocking is encoded by the absence of transition labeled with \styleact{endT1} in location \styleloc{wait\_T3} (an alternative is to synchronize on \styleact{endT1} to a sink location that also blocks time elapsing).
\end{example}

Both remaining reactivities in \cref{figure:reactivities} follow easily from this scheme: the first reactivity (Navigation $\rightarrow$ Guidance $\rightarrow$ Control) follows the same principle for Navigation and Guidance, and is immediately followed by a third check for Control, while the second reactivity (Navigation $\rightarrow$ Control) is simpler as both Navigation and Control are on the same thread.

\subsection{Compositional verification and synthesis}\label{ss:compositional}

Due to the nondeterministic choice, the verification of the reactivities entails a clear overhead to the verification (see \cref{section:experiments}).
Verifying all three reactivities can be naturally done by adding the three observer automata to the same system, and performing synthesis on the composition of all these automata.

However, we claim that this can be done in a \emph{compositional} fashion.
Indeed, checking reactivities is checking that a constraint is met for all executions; this can be seen as a global invariant of the property ``all reactivities are satisfied''%
, and we will verify it using observers.
Observers simply \emph{observe} the system and do not interact with it as long as the property they are verifying is not violated%
; therefore, independent properties can be observed by different observers using different executions.
Therefore, checking that these three invariants are valid can be done separately.
In the non-parametric case, we will perform three different verifications of the system, with only one reactivity automaton at a time.
Then, if the ``bad'' locations are unreachable for the three different verifications, then the system is schedulable and the reactivities are met.
In the case of synthesis, we will \emph{intersect} the result of the synthesis applied to the three parametric models.

This compositional analysis comes in contrast with many works on scheduling, where compositionality is hard to achieve (see, \eg{} \cite{SL03,Richter05,LB05,SEL08,CPV13}).
Note that our compositional verification is not necessarily specific to a \emph{parametric} approach, and using our approach in a non-parametric setting (\eg{} using \uppaal{}) could also benefit from a similar compositionality.

\section{Enhancing the analysis with context switches}\label{section:switch}
\parORsubsection{Problem}

When switching between two threads, the CPU needs to store the state of a thread, so that it can be restored later, and 
		consequently %
	the execution can be resumed from the same point later.
Threads usually do not switch instantaneously: a certain amount of time is required for copying data.
For each change in thread execution, the system must copy data before running the next thread.
The time to save this state and restore another is known as \emph{thread context-switch time}.
This context-switch time between threads is small, but can be important to consider for schedulability.

\begin{example}\label{example:tight}
	For the threads assignment given in \cref{figure:typical_solution} together with the processings values in \cref{figure:example_control_system,figure:example_WCET}, we can show using the \cheddar{} analyzer (see \cref{sss:cheddar}) that the schedule is \emph{tight}, \ie{} the occupancy of the processor is 100\,\%.
	For this reason, any non-zero context-switch time implies that the system becomes non-schedulable.
\end{example}

Because of the tight schedule mentioned in \cref{example:tight}, in order to study the system using non-zero context-switch times, we consider the second set of (fictitious) values, given in \cref{figure:example_WCET_switch_time}.
This set reduces the WCETs of the processings, and therefore allows for some non-zero context-switch time.
\begin{figure}[tb!]
  \centering
\begin{lstlisting}[style=ariane]
processing wcet Navigation (1ms);
processing wcet Guidance   (10.5ms);
processing wcet Control    (3ms);
processing wcet Monitoring (3ms);
\end{lstlisting}
  \caption{Example of Worst Case Execution Times for a system with switch time}
  \label{figure:example_WCET_switch_time}
\end{figure}
\ifdefined\LaTeXdiff\else
\ArXivVersion{
Following the new data from \cref{figure:example_WCET_switch_time},
let us redefine the set of processings as $\Processings' = \{ \procNavi, \procCont, \procMoni, \procGuid \}$, where:

\begin{description}
	\item[Control:] $\procNavi = (\WCETNavi , \PPNavi) = (1, 5) $.

	\item[Guidance:]$\procCont = (\WCETCont , \PPCont) = (3, 10) $.

	\item[Monitoring:] $\procMoni = (\WCETMoni , \PPMoni) = (3, 20) $.

	\item[Navigation:] $\procGuid = (\WCETGuid , \PPGuid) = (10.5, 60) $.
\end{description}
}
\fi

From now on, we consider  that the switch from a thread to another one requires a (constant) switch time equal to 500\,$\mu s$ \footnote{%
	All values are confidential and therefore the given values in this paper are not the genuine ones.%
		\label{footnote:confidential}
	}.
Also note that, during the change of processing within the same thread (\eg{} from Navigation to Control in the odd period of Thread~T1 in \cref{figure:typical_solution}), the switch remains~0, as these processings are part of the \emph{same} thread.

\parORsubsection{Modeling the context switch}

The switch time between threads is modeled as part of the scheduler automaton.
For each change of execution from one thread to another, we go through an intermediate location:
upon activation of a thread implying a thread switch (recall that some thread activations may not imply an immediate thread change, if the newly activated thread has lesser priority than the currently activated thread, \eg{} activating Thread T2 in location \styleloc{execT1} in \cref{figure:scheduler}),
	then a clock \styleparam{xSwitch} is set to~0, and counts until it reaches the context switch time.
In our case, this timing is parameter \styleparam{pSwitch} (this parameter is in practice assigned to its nominal value 500\,$\mu s$ and is therefore not truly parametric).

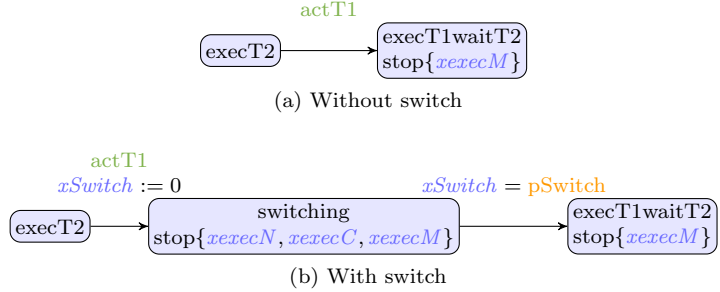
\begin{figure}
	\centering
	\subfloat[Without switch]{
		\begin{tikzpicture}[scale=.75, xscale=1.5, auto, ->, >=stealth']
	
			\node[location] at (0, 0) (execT2) [align=center] {\styleloc{execT2}};
	
			\node[location] at (2.5, 0) (execT1waitT2) [align=center] {\styleloc{execT1waitT2} \\ stop$\{ \styleclock{xexecM} \}$};

			\path (execT2) edge node[align=center]{
			$\styleact{actT1}$\\
			} (execT1waitT2);
		\end{tikzpicture}
		\label{figure:encoding-switch:without}
	}
	\hfill{}
	\subfloat[With switch]{
		\begin{tikzpicture}[scale=.75, xscale=1.5, auto, ->, >=stealth']
	
			\node[location] at (0, 0) (execT2) [align=center] {\styleloc{execT2}};

			\node[location] at (3, 0) (switching) [align=center] {\styleloc{switching}\\ stop$\{ \styleclock{xexecN}, \styleclock{xexecC}, \styleclock{xexecM} \}$};
	
			\node[location] at (7, 0) (execT1waitT2) [align=center] {\styleloc{execT1waitT2} \\ stop$\{ \styleclock{xexecM} \}$};

			\path (execT2) edge node[align=center]{
			$\styleact{actT1}$\\
			$\styleclock{xSwitch} := 0$\\
			} (switching)
			
			(switching) edge node[align=center]{
			$\styleclock{xSwitch} = \styleparam{pSwitch}$\\
			} (execT1waitT2);
		\end{tikzpicture}

		\label{figure:encoding-switch:with}
	}
	\caption{Encoding the \FPS{} scheduler without and with switch (fragment)}
	\label{figure:encoding-switch}
\end{figure}
\begin{example}
	We give in \cref{figure:encoding-switch:without} a fragment of the original \FPS{} scheduler from \cref{figure:scheduler}, corresponding to the execution of thread~T2, followed by the activation of thread T1, which has higher priority and therefore requires a context-switch time.
	In the original version in \cref{figure:encoding-switch:without}, the processor immediately starts executing T1 in location \styleloc{execT1waitT2}.
	In contrast, in the transformed version in \cref{figure:encoding-switch:with}, the processor first transits through an intermediate location \styleloc{switching}, that it can only leave \styleparam{pSwitch} time units later; only from there, T1 starts being executed.
	
	Also note that, in the intermediate location \styleloc{switching}, all clocks measuring the execution times of the processings associated to any of the threads of the processor (here Navigation and Control for~T1 and Monitoring for~T2) are stopped, as the processor is not executing any thread, but is performing the context switch.
\end{example}
\ifdefined\LaTeXdiff\else
\ArXivVersion{
We give in \cref{figure:GNC_with_switch} the Gantt chart of the case study of interest with switch time equal to 500\,$\mu s$.
(This Gantt chart was generated by \cheddar{}~\cite{Cheddar}.)

\begin{figure*}[htbp!]
  \centering
  
  \includegraphics[width=\textwidth]{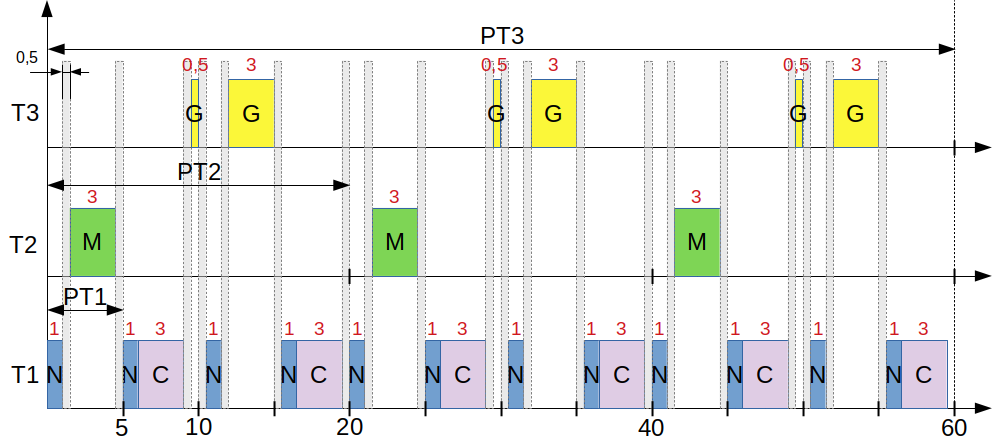}
  \caption{Gantt chart of the system GNC with switch time $= 500\,\mu s$}
  \label{figure:GNC_with_switch}
\end{figure*}
}
\fi

We give in \cref{figure:scheduler-full-version} in \cref{appendix:full-scheduler} the full version of the scheduler with three threads and the switch time between threads.

\section{Experiments}\label{section:experiments}
\subsection{Experimental environment}
We modeled our network of PSA in the \imitator{} input language~\cite{Andre21}.
\imitator{} is a parametric model checker taking as input networks of PSA extended with useful features such as synchronization actions and discrete variables.
Synthesis can be performed using various properties.
We use here reachability synthesis (formalized in, \eg{} \cite{JLR15}).
When \imitator{} terminates (which is not guaranteed in theory), the result is always sound (but not necessarily complete), but the tool is often able to infer whether the result is \emph{exact} (sound and complete).
All analyses mentioned in this manuscript terminate with an exact result.\label{newtext:imitatortermination}

The translation effort was manual due to the specificity of our solution (with the exception of the scheduler, for which we started from an automated generator).
However, we tried to keep our translation as systematic as possible to allow for a future automated generation from the problem input data.
We made intensive use of clock resets and stopwatches for clocks not necessary at some points, in order to let \imitator{} apply inactive clock reductions.

All experiments were conducted using \imitator{} 2.10.4 ``Butter Jellyfish'' on an ASUS X411UN Intel Core\texttrademark{}
i7-8550U 1.80\,GHz %
	with 8\,GiB memory running Linux Mint~19 64\,bits.\footnote{%
	Sources, binaries, models and results are available at \href{https://www.imitator.fr/static/FI2021/}{\nolinkurl{imitator.fr/static/FI2021/}} and at \url{https://www.doi.org/10.5281/zenodo.5042059}.
}

In \cref{ss:XP-without,ss:XP-with}, we first study the system without the extra context switch time introduced in \cref{section:switch};
then, we study in \cref{ss:XP-switch} the overhead incurred by the context switch time.

\subsection{Verification and synthesis without reactivities}\label{ss:XP-without}

\ArXivVersion{
In order to evaluate the overhead of the satisfaction of the reactivities, we first run analyses \emph{without} reactivities.
}

\subsubsection{Non-parametric model}
First, a non-parametric analysis shows that the bad locations are unreachable, and therefore the system is schedulable under the nominal values given in \cref{figure:example_control_system,figure:example_WCET}.
\ifdefined\LaTeXdiff\else\ArXivVersion{

}\fi%
The computation time of this non-parametric analysis, together with other parametric analyses (all without reactivities) are given in \cref{table:time-no-reac}.

\ifdefined\LaTeXdiff\else
\ArXivVersion{
We give in \cref{fig:GNC-with-Cheddar} %
the Gantt chart (obtained with \cheddar{}~\cite{Cheddar}) of this entirely instantiated model.

\begin{figure*}[htbp!]
	\centering
  \includegraphics[width=\textwidth]{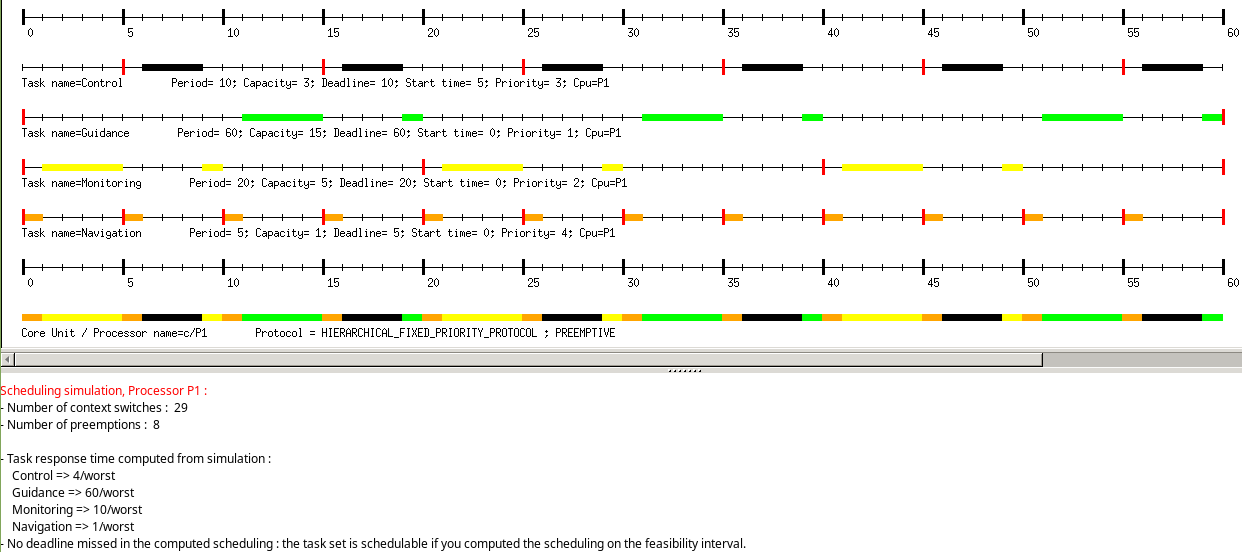}
  \caption{scheduling GNC without reactivities using \cheddar{}}
  \label{fig:GNC-with-Cheddar}
\end{figure*}
}
\fi

\begin{figure}[tb]
	\footnotesize
	\subfloat[Without switch time]{
	\centering
		\begin{minipage}[b]{0.4\textwidth}
			\begin{tabular}{| l | r | }
				\hline
				\cellHeader{Analysis} & \cellHeader{Time (s)} \\
				\hline
				No parameter & 3.1\\
				\hline
				Parametric offsets & 95.8\\
				\hline
				Parametric deadlines & 17.7 \\
				\hline
			\end{tabular}
		\end{minipage}
		\label{table:time-no-reac}
	}
	\hfill
	\subfloat[With switch time]{
	\centering
		\begin{minipage}[b]{0.4\textwidth}
			\begin{tabular}{| l | r | }
				\hline
				\cellHeader{Analysis} & \cellHeader{Time (s)} \\
				\hline
				No parameter & 17.9\\
				\hline
				Parametric offsets & 5,396.3\\
				\hline
				Parametric deadlines & 38.7\\
				\hline
			\end{tabular}
		\end{minipage}
		\label{table:time-no-reac-with-switch}
	}
	\caption{Computation times without reactivities}
\end{figure}

\subsubsection{Parameterized offsets}
We then parameterize offsets, \ie{} we seek admissible offsets for which the system is schedulable.
The constraint synthesized by \imitator{} is given in \cref{table:result-parametric-offsets} in \cref{appendix:parametric-offsets}.
We can see that, while several conditions for schedulability are given, at least one offset must be~0 to ensure schedulability.

In order to exemplify admissible values, we exhibit some valuations satisfying this constraint in \cref{table:no-reac-param-offset-examples} in \cref{appendix:parametric-offsets}; we also give some valuations \emph{not} satisfying this constraint.
These valuations were derived manually from the constraint, but an automatization thanks to an SMT solver would be possible.

\subsubsection{Parameterized deadlines}\label{sss:parametric-deadlines}
We then parameterize deadlines, \ie{} we seek admissible deadlines for which the system is schedulable.
The constraint is:
 $ \deadlineTtwo{} \in [11, 20]
 \land\ \deadlineTone{} \in [4, 5]
 \land\ \deadlineTthree{} = 60$.
That is, the deadline of~T3 is strict, while T1 and~T2 can be relaxed while preserving schedulability.

\ArXivVersion{%
Again, we exhibit some valuations satisfying this constraint in \cref{table:no-reac-param-deadlines-examples} in \cref{appendix:parametric-deadlines}.
}

\subsection{Compositional verification of reactivities}\label{ss:XP-with}

We then solve the scheduling verification and scheduling synthesis problems with reactivities, using two methods:
\begin{oneenumerate}
	\item monolithic verification: all three reactivity automata are included in the model; and
	\item compositional verification: we verify sequentially three different models, each of them including all automata modeling the system, but only one reactivity at a time.
\end{oneenumerate}

\begin{table}[tb]
	\footnotesize
	\centering
	\caption{Computation times with reactivities (s)}
	\begin{tabular}{| l | r | r | r | r | r | }
		\hline
		\cellHeader{Analysis} & \cellHeader{Monolithic}  & \cellHeader{NGC}  & \cellHeader{NC}  & \cellHeader{NM} & \cellHeader{Compositional} \\
		\hline
		No parameter & 109.4 & 21.4 & 3.4 & 15.2 & 40.1 \\
		\hline
		Parametric offsets & 2304.0 & 1111.9 & 210.8 & 955.7  & 2278.4\\
		\hline
		Parametric deadlines & 637.2 & 173.0 & 28.5 & 129.8 & 331.3\\
		\hline
	\end{tabular}
	\label{table:time-reac-full}
\end{table}

We give the various \ArXivVersion{computation }times, including the overhead incurred by each reactivity, in \cref{table:time-reac-full}.
\cref{table:time-reac-full} shows the interest of the compositional verification over monolithic verification, as the computation time is divided by a factor~2, except in the case of parametric offsets, where the compositional verification is just a little more efficient.
Also, without surprise, the most complicated reactivity (NGC) takes the longest computation time.

 \subsection[Switch time]{Switch time}\label{ss:XP-switch}

We now give the computation times in the case of switch time of 500\,$\mu s$ in \cref{table:time-no-reac-with-switch,table:time-reac-full-with-switch}. 
In \cref{table:time-reac-full-with-switch}, ``\cellTO{}'' denotes non-termination after 12~hours.
\begin{table}
	\footnotesize
	\centering
	\caption{Computation times with reactivities and switch time (s)}
	\begin{tabular}{| l | r | r | r | r | r | }
		\hline
		\cellHeader{Analysis} & \cellHeader{Monolithic}  & \cellHeader{NGC}  & \cellHeader{NC}  & \cellHeader{NM} & \cellHeader{Compositional} \\
		\hline
		No parameter & 476.5 & 47.5 & 6.5 & 34.5 & 88.5 \\
		\hline
		Parametric offsets & \cellTO{} & 33,449.6 & 2915.4 & \cellTO  & \cellTO \\
		\hline
		Parametric deadlines & 1,919.7 & 342.3 & 62.7 & 278.0 & 683.0\\
		\hline
	\end{tabular}
	\label{table:time-reac-full-with-switch}
\end{table}

The constraints for offsets and deadlines synthesized by \imitator{} are given respectively in \cref{table:result-parametric-offsets-with-switch-time} in \cref{appendix:parametric-offsets-with-switch-time} and in \cref{table:result-parametric-deadlines-with-switch-time} in \cref{appendix:parametric-deadlines-with-switch-time}.

\begin{table}[tb]
	\centering
	\footnotesize
	\caption{Some valuations for which the system is schedulable (without reactivities)}
	\begin{tabular}{| c | c | c | c | c |}
		\hline
		\cellHeader{Valuation} & \cellHeader{\offsetTone{}} & \cellHeader{\offsetTtwo{}} & \cellHeader{\offsetTthree{}} & \cellHeader{$\models K$}\\
		\hline
		\cellHeader{$\pval_1$} & 0 & 11.5 & 1.5 & \cellYes{} \\
		\hline
		\cellHeader{$\pval_2$} & 3 & 0 & 10 & \cellYes{} \\
		\hline
		\cellHeader{$\pval_3$} & 0 & 5 & 1 & \cellYes{} \\
		\hline
		\cellHeader{$\pval_4$} & 12 & 0 & 3 & \cellNo{} \\
		\hline
		\cellHeader{$\pval_5$} & 3 & 15 & 0 & \cellNo{} \\
		\hline
	\end{tabular}
	\label{table:examples-with-switch-time-offsets}
\end{table}
\begin{table}[tb]
	\centering
	\footnotesize
	\caption{Some valuations for which the system is schedulable (with reactivities)}
	\begin{tabular}{| c | c | c | c | c |}
		\hline
		\cellHeader{Valuation} & \cellHeader{\deadlineTone{}} & \cellHeader{\deadlineTtwo{}} & \cellHeader{\deadlineTthree{}} & \cellHeader{$\models K$}\\
		\hline
		\cellHeader{$\pval_1$} & 4.5 & 20 & 60 & \cellYes{}\\
		\hline
		\cellHeader{$\pval_2$} & 5 & 4.5 & 60 & \cellYes{}\\
		\hline
		\cellHeader{$\pval_3$} & 4.5 & 4.5 & 60 & \cellYes{}\\
		\hline
		\cellHeader{$\pval_4$} & 4 & 5 & 60 & \cellNo{}\\
		\hline
		\cellHeader{$\pval_5$} & 4.5 & 4 & 65 & \cellNo{}\\
		\hline
	\end{tabular}
	\label{table:examples-with-switch-time-deadlines}
\end{table}

We give in \cref{table:examples-with-switch-time-offsets,table:examples-with-switch-time-deadlines} some examples of the values of parameters for which the system with switch time is schedulable (``$\models K$'') or not.

Let us briefly compare the computation times of the system without switch time on the one hand, and of the system with the switch time of 500\,$\mu s$ on the other hand.
The execution time of \imitator{} for the system with the switch time is nearly three times higher than the system without switch time, in the non-parametric case and the case of parametric deadlines.
For the case of parametric offsets, it is nearly ten times higher.

\section{Comparison with other tools}\label{section:comparison}
\subsection{Comparison of our results with non-parametric tools}\label{subsection:comparison}

We perform a comparison with two other well-known tools, one from the real-time system community, namely \cheddar{}~\cite{Cheddar}, and one from the timed automata community, namely \uppaal{}~\cite{LPY97}.
Both tools cannot handle parameters nor consider partially specified problems, and therefore can only solve the scheduling \emph{verification} problem.
Therefore, in this section, we consider the instantiated version of the system according to the nominal values given in \cref{figure:example_control_system,figure:example_WCET}.
In addition, to the best of our knowledge, \cheddar{} cannot test the reactivities.

\subsubsection{Non-parametric comparison with \cheddar{}}\label{sss:cheddar}

\cheddar{} is a real-time scheduling tool distributed under the GPL license.
\cheddar{} is used to model software architectures of real-time systems and to check whether the system is schedulable.

We checked the system's schedulability using \cheddar{} when the system is instantiated (\ie{} all offsets are initialized to 0 and the deadline of each thread equal to the period).
We have indicated the period, the execution time and deadline of each processings.
\ifdefined\LaTeXdiff\else\ArXivVersion{

}\fi%
As result, \cheddar{} proves that the system in \cref{figure:example_WCET} without switch time between threads is schedulable and there is no deadline missed in the computed scheduling.
\ifdefined\LaTeXdiff\else
\ArXivVersion{%
	We give in \cref{fig:GNC-with-Cheddar} the Gantt chart of this system using \cheddar{}.
}\fi%
The computation time of this analysis is given in \cref{table:comparison-no-parameters}.
	In this solution, the number of context switches per period of~T3 is~29 and the number of preemptions is~8.

\cheddar{} cannot give a solution to the scheduling synthesis problem since it only works with instantiated systems, so we cannot determine offsets and deadlines, and also it does not deal with reactivities.

\subsubsection{Non-parametric comparison with \uppaal{}}

We also compare the obtained results using \imitator{} with \uppaal{} results (for the model without switch time).
\uppaal{} is a timed model checker taking as input networks of timed automata, extended with some useful features such as synchronization, integer-valued global variables, data structures and C-style functions.
We wrote a \uppaal{} model identical to the \imitator{} model---with instantiated parameters as \uppaal{} does not support parametric analyses.
\ifdefined\LaTeXdiff\else\ArXivVersion{%

}\fi%
As result, \uppaal{} proves that the instantiated system is schedulable, both without and with reactivities.
\ifdefined\LaTeXdiff\else\ArXivVersion{%
We give in \cref{table:param-offset-with-switch-time-examples-uppaal} obtained results using \uppaal{} when offsets are parameterized  and in \cref{table:param-deadlines-with-switch-time-examples-uppaal} when deadlines are parameterized.

\begin{table}[tb]
	\centering
	\footnotesize
 	\caption{Possible offset valuations with switch time using \uppaal{}}
	\begin{tabular}{| c | c | c | c | c |}
		\hline
		\cellHeader{Valuation} & \cellHeader{\offsetTone{}} & \cellHeader{\offsetTtwo{}} & \cellHeader{\offsetTthree{}} &\cellHeader{\makecell{\uppaal{} \\ (with reactivities)}}\\
		\hline
		\cellHeader{$\pval_1$} & 0 & 11.5 & 1.5 &  \cellYes{}\\
		\hline
		\cellHeader{$\pval_2$} & 3 & 0 & 10 & \cellYes{}\\
		\hline
		\cellHeader{$\pval_3$} & 0 & 5 & 1 & \cellYes{}\\
		\hline
		\cellHeader{$\pval_4$} & 12 & 0 & 3 & \cellNo{}\\
		\hline
		\cellHeader{$\pval_5$} & 3 & 15 & 0 & \cellNo{}\\
		\hline
	\end{tabular}
	\label{table:param-offset-with-switch-time-examples-uppaal}
\end{table}
\begin{table}[tb]
	\footnotesize
	\centering
 	\caption{Possible deadline valuations with switch time using \uppaal{}}
	\begin{tabular}{| c | c | c | c | c |}
		\hline
		\cellHeader{Valuation} & \cellHeader{\deadlineTone{}} & \cellHeader{\deadlineTtwo{}} & \cellHeader{\deadlineTthree{}} & \cellHeader{\makecell{\uppaal{} \\ (with reactivities)}}\\
		\hline
		\cellHeader{$\pval_1$} & 4.5 & 20 & 60 & \cellYes{} \\
		\hline
		\cellHeader{$\pval_2$} & 5 & 4.5 & 60 & \cellYes{}\\
		\hline
		\cellHeader{$\pval_3$} & 4.5 & 4.5 & 60 & \cellYes{}\\
		\hline
		\cellHeader{$\pval_4$} & 4 & 5 & 60 & \cellNo{}\\
		\hline
		\cellHeader{$\pval_5$} & 4.5 & 4 & 65 & \cellNo{}\\
		\hline
	\end{tabular}
	\label{table:param-deadlines-with-switch-time-examples-uppaal}
\end{table}
}\fi%
\ifdefined\LaTeXdiff
	We checked that all valuations for which the system is schedulable (resp.\ not schedulable) in \cref{table:examples-with-switch-time-offsets,table:examples-with-switch-time-deadlines} are indeed proved to be schedulable (resp.\ not schedulable) by \uppaal{}.
\else
\NOTArXivVersion{%
	We checked that all valuations for which the system is schedulable (resp.\ not schedulable) in \cref{table:examples-with-switch-time-offsets,table:examples-with-switch-time-deadlines} are indeed proved to be schedulable (resp.\ not schedulable) by \uppaal{}.
}
\fi

\subsubsection{Summary of comparisons}

We give the computation times without reactivities in \cref{table:comparison-no-parameters}.
Clearly, from our experiments, if the model features no parameters, \cheddar{} (if no reactivities are specified) or \uppaal{} (if some reactivities are specified) should be used.
However, none of these tools cope with uncertain constants.
Therefore, despite the complexity overhead, \imitator{} should be used if some timing constants are unspecified.

\begin{table}
	\centering
	\footnotesize
	\caption{Computation times without parameters}
	\begin{tabular}{| l | r | r | }
		\hline
		\cellHeader{Analysis} & \cellHeader{Without reactivities (s)} & \cellHeader{With reactivities (s)} \\
		\hline
		\cheddar{} & $<0.1$ & \cellNA{}\\
		\hline
		\imitator{} & 3.086 & 109.404 \\
		\hline
		\uppaal{} & 0.002 & 0.003\\
		\hline
	\end{tabular}
	\label{table:comparison-no-parameters}
\end{table}
\subsection{``Testing'' the parametric analysis}\label{ss:verification-of-verification}

Finally, we tried to obtain additional guarantees on our \emph{model's} correctness.
	Indeed, while we can reasonably suppose that our methodology is correct and that the tools are exempt from bugs for the algorithms used here (which remains to be done formally though), a major issue is that of the manual coding of our model into the input language of \imitator{}.
In order to have further guarantees, we compared several aspects of the results with other results, or with other tools, whenever applicable.

\subsubsection{Using non-parametric model checking}

In order to increase our confidence in the results obtained with \imitator{} in \cref{ss:XP-without}, we will first \emph{test} that sampled valuations from the parametric constraint synthesized by \imitator{} are indeed proved schedulable (resp.\ non-schedulable) by non-parametric tools whenever they belong (resp.\ do not belong) to the constraint synthesized by \imitator{}.
Once more, we do so using both a popular tool in the real-time systems community (\cheddar{}) and a non-parametric timed model checker (\uppaal{}).

\paragraph{Model with reactivities}
First, we fix the deadlines, and we vary the offsets according to the constraint synthesized in \cref{ss:XP-with}.
We sample four valuations of this constraint, and give them in \cref{table:param-offset-examples-cheddar-uppaal} ($\pval_1$ to~$\pval_4$);
we also add two valuations ($\pval_5$ and~$\pval_6$) not belonging to the constraint synthesized by \imitator{}.
For each of these valuations, we test using \cheddar{} and \uppaal{} whether the system is schedulable.

\begin{table}[tb]
	\centering
	\scriptsize
 	\caption{Possible offset valuations (with reactivities) checked using \cheddar{} and \uppaal{}}
 	\ifdefined\VersionArXiv
		\setlength{\tabcolsep}{2pt} %
 	\fi
	\begin{tabular}{| c | c | c | c | c | c | c |}
		\hline
		\cellHeader{Valuation} & \cellHeader{\offsetTone{}} & \cellHeader{\offsetTtwo{}} & \cellHeader{\offsetTthree{}} & \cellHeader{$\models K$} & \cellHeader{\makecell{ \cheddar{}\\ (without reactivities)}} &\cellHeader{\makecell{\uppaal{} \\ (with reactivities)}}\\
		\hline
		\cellHeader{$\pval_1$} & 0 & 2 & 1 & \cellYes{} & \cellYes{} & \cellYes{}\\
		\hline
		\cellHeader{$\pval_2$} & 4 & 0 & 10 & \cellYes{} & \cellYes{} & \cellYes{}\\
		\hline
		\cellHeader{$\pval_3$} & 2 & 10 & 0 & \cellYes{} & \cellYes{} & \cellYes{}\\
		\hline
		\cellHeader{$\pval_4$} & 0 & 9 & 0 & \cellYes{} & \cellYes{} & \cellYes{}\\
		\hline
		\cellHeader{$\pval_5$} & 2 & 12 & 1 & \cellNo{} & \cellNo{} & \cellNo{}\\
		\hline
		\cellHeader{$\pval_6$} & 5 & 9 & 0 & \cellNo{} & \cellNo{} & \cellNo{}\\
		\hline
	\end{tabular}
	\label{table:param-offset-examples-cheddar-uppaal}
\end{table}
\begin{table}[tb]
	\centering
	\scriptsize
 	\caption{Possible deadline valuations (with reactivities) checked using \cheddar{} and \uppaal{}}
 	\ifdefined\VersionArXiv
		\setlength{\tabcolsep}{1pt} %
 	\fi
	\begin{tabular}{| c | c | c | c | c | c | c |}
		\hline
		\cellHeader{Valuation} & \cellHeader{\deadlineTone{}} & \cellHeader{\deadlineTtwo{}} & \cellHeader{\deadlineTthree{}} & \cellHeader{$\models K$} & \cellHeader{\makecell{ \cheddar{}\\ (without reactivities)}} &\cellHeader{\makecell{\uppaal{} \\ (with reactivities)}}\\

		\hline
		\cellHeader{$\pval_1$} & 5 & 20 & 60 & \cellYes{} & \cellYes{} & \cellYes{}\\
		\hline
		\cellHeader{$\pval_2$} & 4 & 11 & 60 & \cellYes{} & \cellYes{} & \cellYes{}\\
		\hline
		\cellHeader{$\pval_3$} & 5 & 15 & 60 & \cellYes{} & \cellYes{} & \cellYes{}\\
		\hline
		\cellHeader{$\pval_4$} & 4 & 20 & 60 & \cellYes{} & \cellYes{} & \cellYes{}\\
		\hline
		\cellHeader{$\pval_5$} & 3 & 11 & 60 & \cellNo{} & \cellNo{} & \cellNo{}\\
		\hline
		\cellHeader{$\pval_6$} & 4 & 9 & 55 & \cellNo{} & \cellNo{} & \cellNo{} \\
		\hline
	\end{tabular}
	\label{table:param-deadlines-examples-cheddar-uppaal}
\end{table}

We give in \cref{table:param-offset-examples-cheddar-uppaal} obtained results using \cheddar{} and \uppaal{} when deadlines are instantiated (and offsets remain parameterized) and in \cref{table:param-deadlines-examples-cheddar-uppaal} when offsets are instantiated (and deadlines remain parameterized).
As one can see from \cref{table:param-offset-examples-cheddar-uppaal,table:param-deadlines-examples-cheddar-uppaal}, all results are consistent.
Recall that this does not formally \emph{prove} the correctness of our method, but increases our confidence by \emph{testing sample points}.
Still, if one considers that \uppaal{} or \cheddar{} are reliable tools and that our model is entirely correct, once a given valuation is chosen from the constraint output by \imitator{}, checking again its correctness using one of the aforementioned tools is a good way to assess the validity of the whole process.

\subsubsection{Using constraints comparisons}

We now perform additional tests on the results of \imitator{}.
\ArXivVersion{%
\paragraph{Model without switch time}
}%
We consider here the constraints for the full system including reactivities but excluding the switch time \cref{ss:XP-with} (the case of the switch time gave similar results).

\paragraph{Constraints comparisons}
First, we verified using PolyOp\footnote{%
	A simple interface over the Parma Polyhedra Library~\cite{BHZ08}, available at \href{https://github.com/etienneandre/PolyOp}{\nolinkurl{github.com/etienneandre/PolyOp}}, and that allows for polyhedral computations such as intersection or difference, as well as polyhedral checks such as equality or (strict) inclusion.
}
that the constraint obtained by monolithic verification is equal to the intersection of the 3 constraints (reactivity NC, reactivity NM and reactivity NGC) obtained by separate verifications, on the one hand when offsets are parameterized, and on the other hand when deadlines are parameterized.

Second, we checked that the results with reactivities are included in the constraint without reactivities in all three cases (no parameter, parametric offsets, parametric deadlines).
Indeed, the model with reactivities is more constrained, and therefore its admissible valuations set shall be included in or equal to the valuations set without reactivity constraints.

\paragraph{Constraints difference}

We give %
below the difference of constraints without reactivities and constraints with reactivities when offsets are parameterized:

{\centering
 	\( \offsetTthree{} + 5 > \offsetTtwo{} %
		\land\ \offsetTone{} = 0
		\land\ \offsetTtwo{} \in (1, 5] %
		\land\ \offsetTthree{} \in [0,1]\)
		
}

This shows that the two constraints are not equal: some valuations ensure schedulability when reactivities are not considered, but do not ensure schedulability under reactivity constraints.
\emph{This is a major outcome of our experiments, as it justifies for the analysis under reactivity constraints.}
That is, tools that are not able to test schedulability under reactivity constraints (such as \cheddar{}) will give incorrect results for this case study.
\ifdefined\LaTeXdiff\else\ArXivVersion{

}\fi%
We present in \cref{table:diff-of-without-reacts-with-reacts-examples} some examples of values of offsets in the difference of constraints without and with reactivities: as expected, \cheddar{} mistakenly guarantees the system is correct while \uppaal{} shows it is not, due to some violated reactivities.

\begin{table}[tb]
	\centering
	\footnotesize
 	\caption{Possible offset valuations in the difference of constraints without and with reactivities}
	\begin{tabular}{| c | c | c | c | c | c |}
		\hline
		\cellHeader{Valuation} & \cellHeader{\offsetTone{}} & \cellHeader{\offsetTtwo{}} & \cellHeader{\offsetTthree{}}  & \cellHeader{\makecell{ \cheddar{}}} &\cellHeader{\makecell{\uppaal{}}} \\ 
		\hline
		\cellHeader{$\pval_1$} & 0 & 2 & 0 & \cellYes{} & \cellNo{} \\
		\hline
		\cellHeader{$\pval_2$} & 0 & 4 & 1 & \cellYes{} & \cellNo{}  \\
		\hline
		\cellHeader{$\pval_3$} & 0 & 3 & 1 & \cellYes{} & \cellNo{}  \\
		\hline
	\end{tabular}
	\label{table:diff-of-without-reacts-with-reacts-examples}
\end{table}
\ifdefined\LaTeXdiff\else\ArXivVersion{%
\paragraph{Model with switch time}

We finally perform additional verifications on the results of \cref{ss:XP-switch}.
We verify using PolyOp that the constraint with switch time obtained by monolithic verification is equal to the intersection of the 3 constraints (reactivity NC, reactivity NM and reactivity NGC) obtained by separate verifications when deadlines are parameterized, just as we did in \cref{section:compositional}.
We also checked that the result with reactivity NC is included in the constraint without reactivities when offsets are parameterized; and similarly for
the result with all 3~reactivities.
Not all situations could be considered, as some analyses reach time out (see \cref{table:time-reac-full-with-switch}).

We also rechecked the sample results obtained by \imitator{} in \cref{table:examples-with-switch-time-offsets,table:examples-with-switch-time-deadlines} using \uppaal{}.
We did not use \cheddar{} in this example because, to the best of our knowledge, \cheddar{} cannot apply switch time between threads.

}\fi%
\section{Conclusion and perspectives}\label{section:conclusion}
We proposed an approach to synthesize timing valuations ensuring schedulability of the flight control of a space launcher.
A key issue is to ensure that the system reactivities are met---for which we proposed a compositional solution.

Our implementation of the flight control system as an extension of the parametric timed automata formalism using \imitator{} allows to determine offsets and deadlines of each thread taking into account that all reactivities are satisfied, and allows to ensure formally that the FPS type scheduling can be a solution for our problem. 
We build a modular solution, \ie{} we specified an automaton for each element of our system (thread, processing, scheduling policy) and for each constraint (reactivity).
The interest of using \imitator{} as the main tool of our approach is that it allows to analyze a system with parameters in order to determine the possible values of those parameters, unlike other existing tools (\eg{} \cheddar{} and \uppaal{}) which treat only initialized systems.
In addition, we showed that the reactivity constraints are important as, without them, wrong valuations can be derived.

\paragraph*{Future works}

Due to the efficiency gap\ArXivVersion{ of an order of magnitude}, combining some non-parametric analyses (\eg{} with \uppaal{} or \cheddar{}) with parametric analyses (\imitator{}) would be an interesting future work.

The harmonic periods are a strong assumption of the problem.
Tuning our solution to benefit from this assumption is on our agenda.
This may indeed allow us to \ArXivVersion{ reuse some clocks, and therefore} reduce the number of clocks; it is well-known that the model-checking problem is exponential in the number of clocks.

In addition, \emph{synthesizing} the admissible values for the context switch time, \ie{} making this time a parameter, seems interesting as it derives admissible values of the processor context switch speed for which the system can be scheduled.

So far, our method allows us to prove formally properties for systems that contain only  a limited number of threads and processings. In order to scale up the method, it would be interesting to combine it with Bini-Buttazzo's uniprocessor taskset generation method~\cite{bini2005measuring,Emberson2010a}.\label{newtext:bini}

We envisage two tracks for longer-term future works:
\begin{ienumerate}
	\item Generalizing the flight control scheduling problem by automatically synthesizing the allocations of processings on threads.
		This generalization raises first the issue of modeling such a problem (how to model an allocation with a parameter) and second the classical combinatorial explosion of states.
	\item Applying this approach to the automatic synthesis of the launcher sequential, \ie{} of the scheduling of all the system events necessary to fulfill a mission: ignition and shut-down of stages, release of firing, release of payloads, etc.
\end{ienumerate}

\section*{Acknowledgements}
We would like to thank the anonymous reviewers for useful comments.

	\newcommand{\CCIS}{Communications in Computer and Information Science}
	\newcommand{\ENTCS}{Electronic Notes in Theoretical Computer Science}
	\newcommand{\FMSD}{Formal Methods in System Design}
	\newcommand{\IJFCS}{International Journal of Foundations of Computer Science}
	\newcommand{\IJSSE}{International Journal of Secure Software Engineering}
	\newcommand{\JLAP}{Journal of Logic and Algebraic Programming}
	\newcommand{\JLC}{Journal of Logic and Computation}
	\newcommand{\LMCS}{Logical Methods in Computer Science}
	\newcommand{\LNCS}{Lecture Notes in Computer Science}
	\newcommand{\RESS}{Reliability Engineering \& System Safety}
	\newcommand{\STTT}{International Journal on Software Tools for Technology Transfer}
	\newcommand{\TCS}{Theoretical Computer Science}
	\newcommand{\ToPNoC}{Transactions on Petri Nets and Other Models of Concurrency}
	\newcommand{\TSE}{{IEEE} Transactions on Software Engineering}

\ifdefined\VersionArXiv
	\renewcommand*{\bibfont}{\small}
	\printbibliography[title={References}]
\else
	\bibliographystyle{fundam}
	\bibliography{Ariane}
\fi

\newpage
\appendix
\section{Full scheduler}\label{appendix:full-scheduler}

\begin{figure*}[h!]
  \centering
  
  \includegraphics[width=\textwidth]{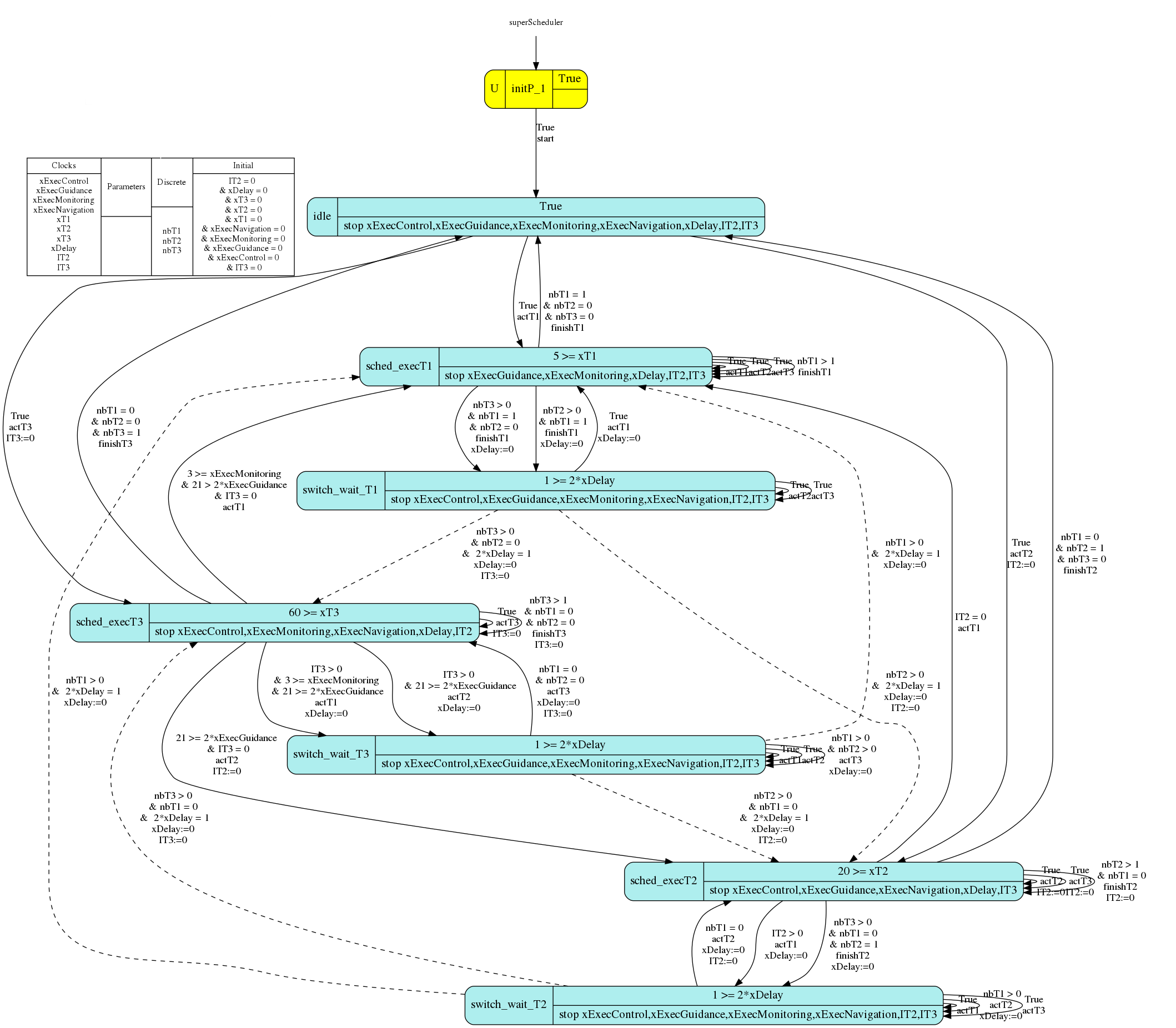}
  \caption{Encoding the \FPS{} scheduler with switches (full version)}
  \label{figure:scheduler-full-version}
\end{figure*}

We give in \cref{figure:scheduler-full-version} the full version of the scheduler with three threads and the switch time between threads.

\section{Parametric analyses without reactivities}
\subsection{Parametric offsets}\label{appendix:parametric-offsets}

The constraint synthesized by \imitator{} for the model with parametric offsets is given in \cref{table:result-parametric-offsets}.

\begin{figure}[htbp!]
	\centering
	\scriptsize
	\begin{tabular}{| l |}
	\hline
$ 5 \geq \offsetTtwo{}$\\
$\land\ \offsetTthree{} + 5 > \offsetTtwo{}$\\
$\land\ \offsetTthree{} \geq 0$\\
$\land\ \offsetTtwo{} \geq 0$\\
$\land\ 1 \geq \offsetTthree{}$\\
$\land\ \offsetTone{} = 0$\\
\textbf{OR}\\
$  \offsetTone{} \geq 0$\\
$\land\ 11 \geq \offsetTthree{}$\\
$\land\ \offsetTthree{} > 1 + \offsetTone{}$\\
$\land\ 4 \geq \offsetTone{}$\\
$\land\ \offsetTtwo{} = 0$\\
\textbf{OR}\\
$  \offsetTthree{} > 1$\\
$\land\ 11 \geq \offsetTthree{}$\\
$\land\ \offsetTtwo{} > 0$\\
$\land\ 1 \geq \offsetTtwo{}$\\
$\land\ \offsetTone{} = 0$\\
\hline
\end{tabular}
\begin{tabular}{| l |}
\hline
\textbf{OR}\\
$  \offsetTone{} > 0$\\
$\land\ \offsetTtwo{} \geq 0$\\
$\land\ 11 \geq \offsetTtwo{}$\\
$\land\ 4 \geq \offsetTone{}$\\
$\land\ \offsetTthree{} = 0$\\
\textbf{OR}\\
$  11 \geq \offsetTtwo{}$\\
$\land\ \offsetTthree{} \geq 0$\\
$\land\ \offsetTtwo{} > 9$\\
$\land\ 1 \geq \offsetTthree{}$\\
$\land\ \offsetTone{} = 0$\\
\textbf{OR}\\
$  \offsetTone{} + 1 \geq \offsetTthree{}$\\
$\land\ \offsetTone{} > 0$\\
$\land\ \offsetTthree{} > 0$\\
$\land\ 4 \geq \offsetTone{}$\\
$\land\ \offsetTtwo{} = 0$\\
\hline
\end{tabular}
\begin{tabular}{| l |}
\hline
\textbf{OR}\\
$  \offsetTtwo{} > 5$\\
$\land\ 9 \geq \offsetTtwo{}$\\
$\land\ \offsetTthree{} > 0$\\
$\land\ 1 \geq \offsetTthree{}$\\
$\land\ \offsetTone{} = 0$\\
\textbf{OR}\\
$  \offsetTtwo{} \geq 5$\\
$\land\ 9 \geq \offsetTtwo{}$\\
$\land\ \offsetTone{} = 0$\\
$\land\ \offsetTthree{} = 0$\\
		\hline
	\end{tabular}
	
	\caption{Parametric offsets}
	\label{table:result-parametric-offsets}
\end{figure}

In order to exemplify admissible values, we exhibit in \cref{table:no-reac-param-offset-examples} some valuations satisfying the constraint in \cref{table:result-parametric-offsets}.

\begin{table}[tb]
	\footnotesize
	\centering
 	\caption{Possible offset valuations (without reactivities)}
	\begin{tabular}{| c | c | c | c | c |}
		\hline
		\cellHeader{Valuation} & \cellHeader{\offsetTone{}} & \cellHeader{\offsetTtwo{}} & \cellHeader{\offsetTthree{}} & \cellHeader{$\models K$}\\
		\hline
		\cellHeader{$\pval_1$} & 0 & 2 & 1 & \cellYes{}\\
		\hline
		\cellHeader{$\pval_2$} & 4 & 0 & 11 & \cellYes{}\\
		\hline
		\cellHeader{$\pval_3$} & 2 & 11 & 0 & \cellYes{}\\
		\hline
		\cellHeader{$\pval_4$} & 0 & 9 & 0 & \cellYes{}\\
		\hline
		\cellHeader{$\pval_5$} & 2 & 12 & 1 & \cellNo{}\\
		\hline
		\cellHeader{$\pval_6$} & 5 & 9 & 0 & \cellNo{}\\
		\hline
	\end{tabular}
	\label{table:no-reac-param-offset-examples}
\end{table}
\ArXivVersion{
\subsection[Parametric deadlines]{Parametric deadlines}\label{appendix:parametric-deadlines}

We exhibit in \cref{table:no-reac-param-deadlines-examples} some valuations satisfying the constraint given in \cref{sss:parametric-deadlines}.	

\begin{table}[tb]
	\footnotesize
	\centering
 	\caption{Possible deadline valuations (without reactivities)}
	\begin{tabular}{| c | c | c | c | c |}
		\hline
		\cellHeader{Valuation} & \cellHeader{\deadlineTone{}} & \cellHeader{\deadlineTtwo{}} & \cellHeader{\deadlineTthree{}} & \cellHeader{$\models K$} \\
		\hline
		\cellHeader{$\pval_1$} & 5 & 20 & 60 & \cellYes{} \\
		\hline
		\cellHeader{$\pval_2$} & 4 & 11 & 60 & \cellYes{} \\
		\hline
		\cellHeader{$\pval_3$} & 5 & 15 & 60 & \cellYes{} \\
		\hline
		\cellHeader{$\pval_4$} & 4 & 20 & 60 & \cellYes{} \\
		\hline
		\cellHeader{$\pval_5$} & 3 & 11 & 60 & \cellNo{} \\
		\hline
		\cellHeader{$\pval_6$} & 4 & 9 & 55 & \cellNo{} \\
		\hline
	\end{tabular}
	\label{table:no-reac-param-deadlines-examples}
\end{table}
}
\subsection{Parametric offsets and deadlines}\label{appendix:parametric-offsets-deadlines}

The constraint synthesized by \imitator{} for the model with both parametric offsets and parametric deadlines is given in \cref{table:result-parametric-offsets-deadlines}.

\begin{figure}[htbp!]
	\centering
	\scriptsize
	
	\ifdefined\VersionArXiv
		\newcommand{\scalefactorHugeTruc}{.8}
	\else
		\newcommand{\scalefactorHugeTruc}{1}
	\fi
	
	\scalebox{\scalefactorHugeTruc}{
	\begin{tabular}{| l |}
	\hline
$\deadlineTtwo{} > 11$\\
$\land\ 11 \geq \offsetTthree{}$\\
$\land\ \deadlineTone{} \geq 4$\\
$\land\ \offsetTthree{} > \offsetTtwo{}$\\
$\land\ 20 \geq \deadlineTtwo{}$\\
$\land\ \offsetTtwo{} > 0$\\
$\land\ 5 \geq \deadlineTone{}$\\
$\land\ 1 \geq \offsetTtwo{}$\\
$\land\ \offsetTone{} = 0$\\
$\land\ \deadlineTthree{} = 60$\\
\textbf{OR}\\
  $\offsetTone{} > 0$\\
$\land\ \deadlineTone{} \geq 4$\\
$\land\ 20 \geq \deadlineTtwo{}$\\
$\land\ \offsetTthree{} > 5$\\
$\land\ \deadlineTtwo{} \geq 15$\\
$\land\ 11 \geq \offsetTthree{}$\\
$\land\ 5 \geq \deadlineTone{}$\\
$\land\ \deadlineTone{} \geq 1 +\offsetTone{}$\\
$\land\ \offsetTtwo{} = 0$\\
$\land\ \deadlineTthree{} = 60$\\
\textbf{OR}\\
  $\offsetTone{} > 0$\\
$\land\ \offsetTone{} + 1 \geq \offsetTthree{}$\\
$\land\ \deadlineTtwo{} \geq 15$\\
$\land\ 20 \geq \deadlineTtwo{}$\\
$\land\ 4 \geq\offsetTone{}$\\
$\land\ \offsetTthree{} \geq 0$\\
$\land\ \deadlineTone{} = 5$\\
$\land\ \offsetTtwo{} = 0$\\
$\land\ \deadlineTthree{} = 60$\\
\textbf{OR}\\
   $\deadlineTtwo{} > 11$\\
$\land\ 11 \geq \offsetTthree{}$\\
$\land\ \deadlineTone{} \geq 4$\\
$\land\ \offsetTthree{} > \deadlineTone{}$\\
$\land\ 20 \geq \deadlineTtwo{}$\\
$\land\ 5 \geq \deadlineTone{}$\\
$\land\ \offsetTtwo{} = 0$\\
$\land\ \offsetTone{} = 0$\\
$\land\ \deadlineTthree{} = 60$\\
\hline
\end{tabular}
\begin{tabular}{| l |}
\hline
\textbf{OR}\\
   $ \deadlineTtwo{} > 11$\\
$\land\ \offsetTthree{} > 0$\\
$\land\ 20 \geq \deadlineTtwo{}$\\
$\land\ 5 \geq \offsetTthree{}$\\
$\land\ \offsetTtwo{} = 0$\\
$\land\ \deadlineTone{} = 5$\\
$\land\ \offsetTone{} = 0$\\
$\land\ \deadlineTthree{} = 60$\\
\textbf{OR}\\
   $20 \geq \deadlineTtwo{}$\\
$\land\ \deadlineTone{} \geq 4$\\
$\land\ \offsetTone{} > 0$\\
$\land\ \deadlineTone{} \geq 1 +\offsetTone{}$\\
$\land\ \deadlineTtwo{} > 11$\\
$\land\ \offsetTthree{} > 1 +\offsetTone{}$\\
$\land\ 5 \geq \deadlineTone{}$\\
$\land\ 5 \geq \offsetTthree{}$\\
$\land\ \offsetTtwo{} = 0$\\
$\land\ \deadlineTthree{} = 60$\\
\textbf{OR}\\
   $\offsetTone{} >= 0$\\
$\land\ \deadlineTtwo{} > 11$\\
$\land\ \offsetTone{} + 1 \geq \offsetTthree{}$\\
$\land\ \deadlineTone{} \geq 4$\\
$\land\ \deadlineTtwo{} \geq 11 +\offsetTone{}$\\
$\land\ 20 \geq \deadlineTtwo{}$\\
$\land\ \offsetTthree{} \geq 0$\\
$\land\ 5 > \deadlineTone{}$\\
$\land\ \deadlineTone{} \geq 1 +\offsetTone{}$\\
$\land\ \offsetTtwo{} = 0$\\
$\land\ \deadlineTthree{} = 60$\\

\textbf{OR}\\
   $\offsetTtwo{} > 5$\\
$\land\ 20 \geq \deadlineTtwo{}$\\
$\land\ \offsetTone{} \geq 0$\\
$\land\ \deadlineTone{} \geq 4$\\
$\land\ \deadlineTtwo{} \geq 10 + \offsetTtwo{}$\\
$\land\ 5 \geq \deadlineTone{}$\\
$\land\ \deadlineTone{} \geq 1 +\offsetTone{}$\\
$\land\ \deadlineTtwo{} \geq 19$\\
$\land\ \offsetTthree{} = 0$\\
$\land\ \deadlineTthree{} = 60$\\
\hline
\end{tabular}
\begin{tabular}{| l |}
\hline
\textbf{OR}\\
 $ \deadlineTtwo{} > 11$\\
$\land\ 1 \geq \offsetTthree{}$\\
$\land\ \deadlineTtwo{} \geq 10 + \offsetTtwo{}$\\
$\land\ \offsetTtwo{} \geq 1$\\
$\land\ 20 \geq \deadlineTtwo{}$\\
$\land\ \deadlineTone{} \geq \offsetTtwo{}$\\
$\land\ 5 > \deadlineTone{}$\\
$\land\ \offsetTthree{} \geq 0$\\
$\land\ \deadlineTone{} \geq 4$\\
$\land\ \offsetTone{} = 0$\\
$\land\ \deadlineTthree{} = 60$\\
\textbf{OR}\\
  $ \deadlineTtwo{} > 11$\\
$\land\ 19 > \deadlineTtwo{}$\\
$\land\ \offsetTtwo{} \geq 1$\\
$\land\ 5 > \offsetTtwo{}$\\
$\land\ \offsetTthree{} > 0$\\
$\land\ \deadlineTtwo{} \geq 10 + \offsetTtwo{}$\\
$\land\ 1 \geq \offsetTthree{}$\\
$\land\ \offsetTone{} = 0$\\
$\land\ \deadlineTone{} = 5$\\
$\land\ \deadlineTthree{} = 60$\\
\textbf{OR}\\
  $ \deadlineTtwo{} > 11$\\
$\land\ \deadlineTone{} \geq 4$\\
$\land\ 5 \geq \deadlineTone{}$\\
$\land\ \offsetTthree{} > 0$\\
$\land\ 1 > \offsetTtwo{}$\\
$\land\ 20 \geq \deadlineTtwo{}$\\
$\land\ \offsetTtwo{} \geq \offsetTthree{}$\\
$\land\ \offsetTone{} = 0$\\
$\land\ \deadlineTthree{} = 60$\\
\textbf{OR}\\
   $19 > \deadlineTtwo{}$\\
$\land\ \offsetTthree{} \geq 0$\\
$\land\ \offsetTtwo{} > \deadlineTone{}$\\
$\land\ \deadlineTtwo{} \geq 10 + \offsetTtwo{}$\\
$\land\ 5 > \offsetTtwo{}$\\
$\land\ 1 \geq \offsetTthree{}$\\
$\land\ \deadlineTone{} \geq 4$\\
$\land\ \offsetTone{} = 0$\\
$\land\ \deadlineTthree{} = 60$\\
		\hline
	\end{tabular}
	
	}

	\caption{Parametric offsets and deadlines}	
	\label{table:result-parametric-offsets-deadlines}
\end{figure}

\section[Parametric analyses with reactivities]{Parametric analyses with reactivities}
\subsection[Parametric offsets]{Parametric offsets}\label{appendix:parametric-offsets-with-reactivities}

The constraint synthesized by \imitator{} for the model with the 3 reactivities and parametric offsets is given in \cref{table:result-parametric-offsets-with-reactivities}.
\begin{figure}[htbp!]
	\centering
	\scriptsize
	\begin{tabular}{| l |}
	\hline
 $11 >= \offsetTtwo{}$\\
$\land\  \offsetTthree{} >= 0$\\
$\land\  \offsetTtwo{} >= \offsetTthree{}$\\
$\land\  1 >= \offsetTthree{}$\\
$\land\  \offsetTone{} = 0$\\
\textbf{OR}\\
 $ \offsetTthree{} > \offsetTtwo{}$\\
$\land\  1 >= \offsetTtwo{}$\\
$\land\  \offsetTtwo{} >= 0$\\
$\land\  11 >= \offsetTthree{}$\\
$\land\  \offsetTone{} = 0$\\
\hline
\end{tabular}
\begin{tabular}{| l |}
\hline
\textbf{OR}\\
 $ \offsetTtwo{} >= 0$\\
$\land\  \offsetTone{} > 0$\\
$\land\  11 >= \offsetTtwo{}$\\
$\land\  4 >= \offsetTone{}$\\
$\land\  \offsetTthree{} = 0$\\
\textbf{OR}\\
 $ \offsetTone{} > 0$\\
$\land\ 11 >= \offsetTthree{}$\\
$\land\ \offsetTthree{} > 0$\\
$\land\  4 >= \offsetTone{}$\\
$\land\ \offsetTtwo{} = 0$\\
		\hline
	\end{tabular}

	\caption{Parametric offsets for model with the 3 reactivities}
	\label{table:result-parametric-offsets-with-reactivities}
\end{figure}
\subsection[Parametric deadlines]{Parametric deadlines}\label{appendix:parametric-deadlines-with-reactivities}

The constraint synthesized by \imitator{} for the model with the 3 reactivities and parametric deadlines in \cref{table:result-parametric-deadlines-with-reactivities}.

\begin{figure}[htbp!]
	\centering
	\scriptsize
	\begin{tabular}{| l |}
	\hline
$ \deadlineTtwo{} >= 11$\\
$\land\  \deadlineTone{} >= 4$\\
$\land\  5 >= \deadlineTone{}$\\
$\land\  20 >= \deadlineTtwo{}$\\
$\land\  \deadlineTthree{} = 60$\\
		\hline
	\end{tabular}
	
	\caption{Parametric deadlines for model with the 3 reactivities}
	\label{table:result-parametric-deadlines-with-reactivities}
\end{figure}

\section{Parametric analyses with switch time without reactivities}
\subsection[Parametric offsets]{Parametric offsets}\label{appendix:parametric-offsets-with-switch-time}

The constraint synthesized by \imitator{} for the model with parametric offsets and the context switch time is given in \cref{table:result-parametric-offsets-with-switch-time}.

\begin{figure}[htbp!]
\centering
	\scriptsize
	\begin{tabular}{| l |}
	\hline
$ \offsetTthree{} \geq 10$\\
$\land\ 7 \geq 2*\offsetTone{}$\\
$\land\ 2*\offsetTone{} > 5 + 2*\offsetTtwo{}$\\
$\land\ 23 \geq 2*\offsetTthree{}$\\
$\land\ \offsetTone{} \geq 3$\\
$\land\ 2*\offsetTtwo{} + 7 > 2*\offsetTone{}$\\
$\land\ \offsetTtwo{} \geq 0$\\
$\land\ \offsetTone{} > 2 + \offsetTtwo{}$\\
\textbf{OR}\\
$  \offsetTtwo{} + 2*\offsetTthree{} + 3 > 3*\offsetTone{}$\\
$\land\ \offsetTtwo{} \geq 0$\\
$\land\ 2*\offsetTthree{} > 1 + 2*\offsetTone{}$\\
$\land\ 19 \geq 2*\offsetTthree{}$\\
$\land\ 2*\offsetTtwo{} + 7 > 2*\offsetTone{}$\\
$\land\ \offsetTone{} \geq 3 + \offsetTtwo{}$\\
$\land\ 7 \geq 2*\offsetTone{}$\\
\textbf{OR}\\
$  \offsetTthree{} > 3$\\
$\land\ \offsetTthree{} \geq 3 + \offsetTtwo{}$\\
$\land\ 2*\offsetTthree{} > 5 + 2*\offsetTtwo{}$\\
$\land\ \offsetTone{} + \offsetTthree{} > 6 + 2*\offsetTtwo{}$\\
$\land\ \offsetTone{} \geq 3 + \offsetTtwo{}$\\
$\land\ \offsetTtwo{} \geq 0$\\
$\land\ 2*\offsetTone{} + 1 \geq 2*\offsetTthree{}$\\
$\land\ 7 \geq 2*\offsetTone{}$\\
\hline
\end{tabular}
\begin{tabular}{| l |}
\hline
\textbf{OR}\\
$  19 \geq 2*\offsetTthree{}$\\
$\land\ \offsetTone{} \geq 0$\\
$\land\ \offsetTthree{} > 5$\\
$\land\ \offsetTtwo{} > 1 + \offsetTone{}$\\
$\land\ 2 > \offsetTtwo{}$\\

\textbf{OR}\\
$  \offsetTone{} \geq 0$\\
$\land\ \offsetTtwo{} > 0$\\
$\land\ \offsetTthree{} > 5 + \offsetTtwo{}$\\
$\land\ \offsetTtwo{} \geq \offsetTone{}$\\
$\land\ \offsetTone{} + 1 \geq \offsetTtwo{}$\\
$\land\ 1 > 2*\offsetTone{}$\\
$\land\ 19 \geq 2*\offsetTthree{}$\\
\textbf{OR}\\
$  \offsetTtwo{} \geq 5$\\
$\land\ 23 \geq 2*\offsetTtwo{}$\\
$\land\ \offsetTone{} \geq 0$\\
$\land\ \offsetTthree{} > 1 + \offsetTone{}$\\
$\land\ 3 \geq 2*\offsetTthree{}$\\
\hline
\end{tabular}
\begin{tabular}{| l |}
\hline
\textbf{OR}\\
$  23 \geq 2*\offsetTtwo{}$\\
$\land\ \offsetTthree{} \geq 0$\\
$\land\ \offsetTtwo{} \geq 5$\\
$\land\ 1 \geq \offsetTthree{}$\\
$\land\ \offsetTone{} = 0$\\
\textbf{OR}\\
$  2*\offsetTtwo{} > 7$\\
$\land\ 3*\offsetTtwo{} > 4$\\
$\land\ \offsetTtwo{} + 8 > 0$\\
$\land\ 10 > \offsetTtwo{}$\\
$\land\ 2*\offsetTone{} = 7$\\
$\land\ \offsetTthree{} = 0$\\
		\hline
	\end{tabular}

	\caption{Parametric offsets for model with switch time}
	\label{table:result-parametric-offsets-with-switch-time}
\end{figure}

\subsection[Parametric deadlines]{Parametric deadlines}\label{appendix:parametric-deadlines-with-switch-time}

The constraint synthesized by \imitator{} for the model with parametric deadlines and the context switch time is given in \cref{table:result-parametric-deadlines-with-switch-time}.

\begin{figure}[htbp!]
	\centering
	\scriptsize
	\begin{tabular}{| l |}
	\hline
	
$2*\deadlineTtwo{} \geq 9$\\
$\land\ 2*\deadlineTone{} \geq 9$\\
$\land\ 5 \geq \deadlineTone{}$\\
$\land\ 20 \geq \deadlineTtwo{}$\\
$\land\ \deadlineTthree{} = 60$\\
		\hline
	\end{tabular}
	
	\caption{Parametric deadlines for model with switch time}
	\label{table:result-parametric-deadlines-with-switch-time}
\end{figure}

\section[Parametric analyses with reactivities and with switch time]{Parametric analyses with reactivities and with switch time}
\subsection[Parametric offsets]{Parametric offsets}\label{appendix:parametric-offsets-with-switch-time-reactivities}

The constraint synthesized by \imitator{} for the model with parametric offsets, reactivities constraints the context switch time is given in \cref{table:result-parametric-offsets-with-switch-time-reactivities}.

\begin{figure}[htbp!]
\centering
	\tiny
	\scalebox{.98}{
	\begin{tabular}{| l |}
	\hline
$   \offsetTthree{} >= 10$\\
$\land\ 7 >= 2*\offsetTone{}$\\
$\land\ 2*\offsetTone{} > 5 + 2*\offsetTtwo{}$\\
$\land\ 23 >= 2*\offsetTthree{}$\\
$\land\ \offsetTone{} >= 3$\\
$\land\ 2*\offsetTtwo{} + 7 > 2*\offsetTone{}$\\
$\land\ \offsetTtwo{} >= 0$\\
$\land\ \offsetTone{} > 2 + \offsetTtwo{}$\\
\textbf{OR}\\
 $  \offsetTtwo{} + 2*\offsetTthree{} + 3 > 3*\offsetTone{}$\\
$\land\ \offsetTtwo{} >= 0$\\
$\land\ 2*\offsetTthree{} > 1 + 2*\offsetTone{}$\\
$\land\ 19 >= 2*\offsetTthree{}$\\
$\land\ 2*\offsetTtwo{} + 7 > 2*\offsetTone{}$\\
$\land\ \offsetTone{} >= 3 + \offsetTtwo{}$\\
$\land\ 7 >= 2*\offsetTone{}$\\
\textbf{OR}\\
$   \offsetTtwo{} >= 5
	\land\ \offsetTthree{} >= 0$\\
$\land\ 23 >= 2*\offsetTtwo{}$\\
$\land\ 1 >= \offsetTthree{}
	\land\ \offsetTone{} = 0$\\
\textbf{OR}\\
 $  \offsetTone{} >= 0
	\land\ \offsetTtwo{} > 0$\\
$\land\ \offsetTthree{} > 5 + \offsetTtwo{}$\\
$\land\ \offsetTtwo{} >= \offsetTone{}$\\
$\land\ \offsetTone{} + 1 >= \offsetTtwo{}$\\
$\land\ 1 > 2*\offsetTone{}$\\
$\land\ 19 >= 2*\offsetTthree{}$\\
\textbf{OR}\\
$   19 >= 2*\offsetTthree{}$\\
$\land\ \offsetTone{} >= 0
	\land\ \offsetTthree{} > 5$\\
$\land\ \offsetTtwo{} > 1 + \offsetTone{}$\\
$\land\ 2 > \offsetTtwo{}$\\
\textbf{OR}\\
 $  23 >= 2*\offsetTtwo{}$\\
$\land\ \offsetTtwo{} >= 5
	\land\ \offsetTone{} >= 0$\\
$\land\ \offsetTthree{} > 1 + \offsetTone{}$\\
$\land\ 3 >= 2*\offsetTthree{}$\\
\textbf{OR}\\
$   2*\offsetTthree{} + 1 > 2*\offsetTone{} + 2*\offsetTtwo{}$\\
$\land\ \offsetTthree{} >= 3 + \offsetTtwo{}$\\
$\land\ 2*\offsetTthree{} > 5 + 2*\offsetTtwo{}$\\
$\land\ \offsetTone{} + \offsetTthree{} > 6 + 2*\offsetTtwo{}$\\
$\land\ \offsetTone{} >= 3 + \offsetTtwo{}$\\
$\land\ 2*\offsetTone{} + 1 >= 2*\offsetTthree{}$\\
$\land\ 7 >= 2*\offsetTone{}$\\
$\land\ \offsetTtwo{} >= 0$\\
\textbf{OR}\\
$   2*\offsetTone{} > 7$\\
$\land\ \offsetTtwo{} >= 0
	\land\ \offsetTthree{} > 5$\\
$\land\ \offsetTone{} > 3 + \offsetTtwo{}$\\
$\land\ 19 >= 2*\offsetTthree{}
	\land\ 4 >= \offsetTone{}$\\
\textbf{OR}\\
$   \offsetTtwo{} >= 0$\\
$\land\ \offsetTone{} >= 3 + \offsetTtwo{}$\\
$\land\ \offsetTthree{} >= \offsetTtwo{}$\\
$\land\ 4*\offsetTtwo{} + 7 > 2*\offsetTone{}$\\
$\land\ \offsetTtwo{} + 3 > \offsetTthree{}$\\
$\land\ 2*\offsetTtwo{} + 7 >= 2*\offsetTone{}$\\
$\land\ 1 >= 2*\offsetTtwo{}$\\
\textbf{OR}\\
$   2*\offsetTtwo{} > 7$\\
$\land\ 68*\offsetTtwo{} > 13$\\
$\land\ 6*\offsetTtwo{} + 43 > 0$\\
$\land\ 5*\offsetTtwo{} + 52 > 0$\\
$\land\ 10 > \offsetTtwo{}$\\
$\land\ 2*\offsetTone{} = 7
	\land\ \offsetTthree{} = 0$\\
\hline
\end{tabular}
\begin{tabular}{| l |}
\hline
\textbf{OR}\\
$   \offsetTthree{} > 11$\\
$\land\ \offsetTone{} + 1 >= \offsetTtwo{}$\\
$\land\ \offsetTone{} >= 0$\\
$\land\ 1 > 2*\offsetTone{}$\\
$\land\ \offsetTtwo{} > 0$\\
$\land\ \offsetTtwo{} >= \offsetTone{}$\\
$\land\ 23 >= 2*\offsetTthree{}$\\
\textbf{OR}\\
$   1 > 2*\offsetTone{}$\\
$\land\ \offsetTthree{} > 5$\\
$\land\ \offsetTone{} + 1 >= \offsetTtwo{}$\\
$\land\ \offsetTtwo{} > \offsetTone{}$\\
$\land\ 2*\offsetTtwo{} > 1 + 2*\offsetTone{}$\\
$\land\ \offsetTtwo{} + 5 >= \offsetTthree{}$\\
$\land\ \offsetTone{} >= 0$\\
\textbf{OR}\\
$   \offsetTone{} >= 0$\\
$\land\ 2*\offsetTtwo{} + 19 >= 2*\offsetTthree{}$\\
$\land\ 6*\offsetTtwo{} + 35 > 4*\offsetTthree{}$\\
$\land\ \offsetTthree{} >= 10$\\
$\land\ 2*\offsetTtwo{} >= 3$\\
$\land\ 2 > \offsetTtwo{}$\\
$\land\ \offsetTtwo{} > 1 + \offsetTone{}$\\
\textbf{OR}\\
$   5 >= \offsetTthree{}$\\
$\land\ 9 > \offsetTone{} + \offsetTthree{}$\\
$\land\ 5 > \offsetTone{}$\\
$\land\ \offsetTthree{} > \offsetTone{}$\\
$\land\ 2*\offsetTthree{} > 1 + 2*\offsetTone{}$\\
$\land\ 4 >= \offsetTone{}$\\
$\land\ 2*\offsetTone{} >= 7 + 2*\offsetTtwo{}$\\
$\land\ \offsetTtwo{} >= 0$\\
\textbf{OR}\\
$   3 >= 2*\offsetTtwo{}$\\
$\land\ \offsetTthree{} >= 3 + \offsetTtwo{}$\\
$\land\ \offsetTtwo{} > 0$\\
$\land\ 5 > \offsetTthree{}
	\land\ \offsetTone{} = 0$\\

\textbf{OR}\\
$   \offsetTtwo{} + 3 >= \offsetTthree{}$\\
$\land\ 2*\offsetTone{} + \offsetTtwo{} + 3 > \offsetTthree{}$\\
$\land\ \offsetTthree{} >= \offsetTtwo{}$\\
$\land\ \offsetTone{} >= 0
	\land\ \offsetTtwo{} > 1 + \offsetTone{}$\\
$\land\ 3 >= 2*\offsetTtwo{}$\\
\textbf{OR}\\
$   4*\offsetTtwo{} + 5 > 2*\offsetTone{} + 2*\offsetTthree{}$\\
$\land\ 2*\offsetTone{} + 2*\offsetTtwo{} + 2 > \offsetTthree{}$\\
$\land\ 2*\offsetTtwo{} + 2 > \offsetTthree{}$\\
$\land\ 4*\offsetTone{} + 2*\offsetTtwo{} > 3$\\
$\land\ 2*\offsetTtwo{} >= 3
	\land\ \offsetTone{} >= 0$\\
$\land\ \offsetTthree{} > 3 + \offsetTtwo{}$\\
$\land\ \offsetTtwo{} > 1 + \offsetTone{}$\\
$\land\ 5 >= \offsetTthree{}$\\
\textbf{OR}\\
$   \offsetTthree{} >= 10$\\
$\land\ 2*\offsetTone{} >= 7 + 2*\offsetTtwo{}$\\
$\land\ 23 >= 2*\offsetTthree{}$\\
$\land\ 4 >= \offsetTone{}$\\
$\land\ \offsetTtwo{} >= 0$\\
\textbf{OR}\\
$   \offsetTthree{} >= 10$\\
$\land\ \offsetTtwo{} >= 1 + \offsetTone{}$\\
$\land\ 2 > \offsetTone{} + \offsetTtwo{}$\\
$\land\ 3 > 2*\offsetTtwo{}$\\
$\land\ 2*\offsetTtwo{} + 19 >= 2*\offsetTthree{}$\\
$\land\ \offsetTone{} >= 0$\\
\hline
\end{tabular}
}

\caption{Parametric offsets for model with reactivities and with switch time}
	\label{table:result-parametric-offsets-with-switch-time-reactivities}
\end{figure}

\subsection[Parametric deadlines]{Parametric deadlines}\label{appendix:parametric-deadlines-with-switch-time-reactivities}

The constraint synthesized by \imitator{} for the model with parametric deadlines, reactivities constraints and the context switch time is given in \cref{table:result-parametric-deadlines-with-switch-time-reactivities}.

\begin{figure}[ht!]
	\centering
	\scriptsize
	\begin{tabular}{| l |}
	\hline
	$ 2*\deadlineTtwo{} >= 9$\\
	$\land\ 2*\deadlineTone{} >= 9$\\
	$\land\ 5 >= \deadlineTone{}$\\
	$\land\ 20 >= \deadlineTtwo{}$\\
	$\land\ \deadlineTthree{} = 60$\\
		\hline
	\end{tabular}
	\caption{Parametric deadlines for model with reactivities and with switch time}
	\label{table:result-parametric-deadlines-with-switch-time-reactivities}
\end{figure}

\end{document}